\documentclass[11pt,a4paper,english,twoside]{article}

\usepackage{a4wide}
\usepackage{amssymb, amsmath}
\usepackage{graphicx}
\usepackage[all]{xy}
\usepackage{hyperref}
\hypersetup{linktocpage}
\usepackage{enumerate}
\usepackage{xcolor}
\usepackage{dsfont}
\usepackage{empheq}
\usepackage{cite}
\usepackage{float}

\usepackage{soul}

\newcommand{\beq}{\begin{equation}}
\newcommand{\eeq}{\end{equation}}
\def\bea#1\eea{\begin{align}#1\end{align}}
\def\beal#1\eeal{\begin{subequations}\begin{align}#1\end{align}\end{subequations}}
\newcommand{\nn}{\nonumber}
\newcommand{\w}{\wedge}

\renewcommand{\i}{\ensuremath{\textnormal{i}}}

\def\del {\partial}
\def\d {{\rm d}}
\def\mmm {\mathcal{M}}
\def\tV {\tilde{V}}

\begin{document}
\numberwithin{equation}{section}

\begin{titlepage}

\begin{flushright}
CERN-TH-2018-173
\end{flushright}

\begin{center}

\phantom{DRAFT}

\vspace{2.8cm}

{\LARGE \bf{New constraints on classical de Sitter:\vspace{0.3cm}\\ flirting with the swampland}}\\

\vspace{2 cm} {\Large David Andriot}\\
 \vspace{0.9 cm} {\small\slshape Theoretical Physics Department, CERN\\
1211 Geneva 23, Switzerland}\\
\vspace{0.5cm} {\upshape\ttfamily david.andriot@cern.ch}\\

\vspace{3cm}

{\bf Abstract}
\vspace{0.1cm}
\end{center}

\begin{quotation}
\noindent
We study the existence and stability of classical de Sitter solutions of type II supergravities with parallel $D_p$-branes and orientifold $O_p$-planes. Together with the dilaton and volume scalar fields, we consider a third one that distinguishes between parallel and transverse directions to the $D_p/O_p$. We derive the complete scalar potential for these three fields. This formalism allows us to reproduce known constraints obtained in 10d, and to derive new ones. Specifying to group manifolds with constant fluxes, we exclude a large region of parameter space, forbidding de Sitter solutions on nilmanifolds, semi-simple group manifolds, and some solvmanifolds (at least in some standard algebra basis). In the small remaining region, we identify a stability island, where the three scalars could be stabilized in any de Sitter solution. We discuss these results in the swampland context.
\end{quotation}

\end{titlepage}

\newpage

\tableofcontents

\section{Introduction}

The precision of cosmological observations has dramatically improved in the last decade, such that important constraints have now been set on the variety of existing models. Further constraints are to be expected in the coming years, so it becomes pressing to clarify the situation on the theory side. A question that gets more and more attention is that of knowing which four-dimensional (4d) cosmological model, describing the early universe, can be derived as a low energy effective theory from a quantum gravity theory. Since cosmology typically requires a quantum gravity completion, answering this question would provide a very interesting manner to distinguish between 4d models. The 4d theories that fail to have such a high energy completion are said to lie in the swampland. Various conjectures or criteria have been proposed to determine whether a theory is part of the swampland.

The focus of the present work is on the existence and stability of de Sitter solutions, meaning solutions of a theory with a 4d de Sitter space-time. Given the positivity of the presently measured cosmological constant, many cosmological scenarios include at some point a (meta)stable de Sitter solution, i.e.~a local minimum (with positive value) in a potential describing the evolution of the universe. It is for instance the case of many (post)inflation models, which admit a de Sitter minimum after inflation to allow for the reheating process. Our present universe can also be viewed as attracted towards such a future de Sitter point. Having a (meta)stable de Sitter solution in a 4d cosmological model is therefore not strictly speaking necessary, but it is at least a ``simple'' option, common to many models. Asking whether a quantum gravity theory, here string theory, can consistently admit such a stable de Sitter point as a solution is then a way to test many models with respect to the swampland. A negative answer to this question has been proposed recently, precisely as a new swampland criterion \cite{Obied:2018sgi} (see also \cite{Freivogel:2016qwc, Brennan:2017rbf}). As a consequence of this criterion, more intricate cosmological models are favored \cite{Agrawal:2018own, Banerjee:2018qey, Achucarro:2018vey, Lehners:2018vgi, Kehagias:2018uem, Denef:2018etk, Colgain:2018wgk}.

In this work we focus on so-called classical de Sitter solutions. Those are (perturbative) backgrounds of string theory, with a 10d space-time split as a 4d de Sitter space-time times a 6d compact manifold $\mmm$. These solutions are typically studied as solutions of ten-dimensional type II supergravities with $D_p$-branes and orientifold $O_p$-planes, without any further ingredient. Different stringy approaches to the problem of de Sitter solutions exist: those allow for various corrections or other contributions, and we refer to the recent review \cite{Danielsson:2018ztv} or the introduction of \cite{Andriot:2017jhf} for a list and discussion of those. The advantage of the classical 10d solutions is that deriving an effective 4d model from them is a priori a well-defined procedure; in other words, the connection between a 4d model and the 10d stringy completion is in that case well-controlled, or at least simpler compared to the other approaches. An additional advantage is that the (classical) stability can guarantee that quantum (here $g_s$ and $\alpha'$) corrections are subleading, provided that the corresponding scalar fields are stabilized at appropriate values. The problem with these 10d metastable classical de Sitter solutions is that no example is known, up-to-date; this goes in the direction of the swampland criterion \cite{Obied:2018sgi}. Rather these solutions are subject to many no-go theorems and constraints; in particular, they have been completely excluded from heterotic string \cite{Green:2011cn, Gautason:2012tb, Kutasov:2015eba, Quigley:2015jia}. In view of cosmological and swampland motivations,\footnote{Beyond the implications for cosmology, it is generally an important question to know whether such string backgrounds exist. Those are of interest in different contexts, such as holography \cite{Hull:1998vg, Strominger:2001pn, Balasubramanian:2001nb} in relation with higher spin theories (see e.g.~\cite{Hertog:2017ymy}), as well as for further ideas such as \cite{Verlinde:2016toy}.} it is important to determine whether obtaining such solutions is completely excluded (here in type II supergravities), or whether an example can be found in a very specific corner of the parameter space. It is precisely the aim of this paper to find new and tight constraints, and reduce this way the allowed region in parameter space.

To understand better the situation, it is important to distinguish the existence of solutions and their stability: this provides two different sets of conditions. In type II supergravities, an important number of works have studied one or the other aspect \cite{Maldacena:2000mw, Hertzberg:2007wc, Silverstein:2007ac, Covi:2008ea, Haque:2008jz, Caviezel:2008tf, Flauger:2008ad, Danielsson:2009ff, deCarlos:2009fq, Caviezel:2009tu, Wrase:2010ew, Danielsson:2010bc, Blaback:2010sj, Danielsson:2011au, Shiu:2011zt, Burgess:2011rv, VanRiet:2011yc, Danielsson:2012et, Gautason:2013zw, Kallosh:2014oja, Junghans:2016uvg, Andriot:2016xvq, Junghans:2016abx, Andriot:2017jhf}, and few solutions have been found \cite{Caviezel:2008tf, Flauger:2008ad, Danielsson:2009ff, Caviezel:2009tu, Danielsson:2010bc, Danielsson:2011au}, but they are all unstable. We should stress that these existing solutions all have intersecting $D_p/O_p$. This means that they are parallel along some directions and orthogonal along some others. In particular we always take them to fill the 3d extended space, to preserve 4d Lorentz invariance, so we only consider $p\geq 3$. If any, the orthogonal directions are then necessarily among the remaining compact 6d. Most of the known solutions \cite{Danielsson:2011au} have intersecting $O_6$, and one of them \cite{Caviezel:2009tu} has $O_5$ and $O_7$.

A particular case of the above framework is that of parallel $D_p/O_p$ with $\mmm$ being a group manifold, and restricting to a background with constant fluxes: we will specify to such a setup later in this work. In that case, a 4d effective theory, obtained as a consistent truncation, is expected to be an ${\cal N}=4$ gauged supergravity. De Sitter solutions have been obtained in 4d ${\cal N}=4$ gauged supergravities with specific gaugings, see e.g.~\cite{deRoo:2002jf} and references therein. Considering the theories or gaugings that are obtained by a dimensional reduction on a compact $\mmm$ with an orientifold, as described in \cite{DallAgata:2009wsi, Dibitetto:2010rg}, is however much more restrictive, and no example of de Sitter solution is known in such a 4d ${\cal N}=4$ gauged supergravity. A strong result was obtained on this matter in \cite{Dibitetto:2010rg}: it was shown that gauging algebras in which de Sitter solutions were previously found, namely those being a direct product of semi-simple algebras, are not among the gaugings allowed in the more restricted 4d ${\cal N}=4$ gauged supergravities with compactification origin. This confirms the absence of known example, but does not provide a full no-go theorem against de Sitter solutions in such a setup, and we are not aware of stronger results on this matter (see however \cite{Dibitetto:2011gm} for more constraints). Obtaining such a no-go theorem is one motivation of the present work.

\subsection*{Constraints on classical de Sitter solutions}

To determine regions of parameter space where one can hope to find stable de Sitter solutions, two related approaches have been followed. The first one is to work at 10d with equations of motion and Bianchi identities, and the second one is to work with a corresponding 4d scalar potential, extremize and minimize it. For the latter, a popular approach following \cite{Hertzberg:2007wc, Silverstein:2007ac} has been to consider two scalar fields, the 4d dilaton $\tau$ and 6d volume $\rho$, and study the potential $V(\rho, \tau)$. As we verify explicitly in this work, requiring a de Sitter extremum and satisfying the two extrema equations is equivalent to three 10d equations of motion (e.o.m.)
\beq
\begin{array}{r|c|l}
{\cal R}_4 = \dots >0 & & \mbox{trace of Einstein eq. along 4d} > 0\\
\del_{\tau} V|_0=0 & \longleftrightarrow & \mbox{10d dilaton e.o.m.} \\
\del_{\rho} V|_0=0 & & \mbox{trace of Einstein eq. along 10d or 6d}
\end{array}
\ , \label{3eq}
\eeq
hence the equivalence of the results obtained by either approach. Thanks to these three equations, many constraints on the existence of solutions with {\it parallel} $D_p/O_p$ of {\it single size $p$} were obtained, summarized in Table \ref{tab:before} (see \cite{Wrase:2010ew, Shiu:2011zt} for a summary in 4d, and section 3 of \cite{Andriot:2016xvq} for the same results in 10d; these results do not apply to F-theory). In particular, some non-zero RR fluxes $F_q$ or the NSNS $H$-flux were identified as necessary ingredients to obtain such a solution, otherwise a no-go theorem would apply. One recovers this way the need \cite{Maldacena:2000mw} for a source contribution $T_{10}>0$ \eqref{R4T10F}, possible thanks to the presence of orientifolds. We also recall that the results presented here do not consider the possibility of having $\overline{D}_p$ or $\overline{O}_p$.

After this, another equation has been considered: the Bianchi identity for the RR flux that is sourced by the $D_p/O_p$. From a 4d perspective, this equation is not obtained from the potential and comes as an extra requirement on the theory. This can be interpreted as quantum gravity requirement, in the sense that it descends purely from the 10d original theory; we come back to this idea in Section \ref{sec:ccl}. Adding this sourced Bianchi identity provides new constraints: as shown in full generality in 10d \cite{Blaback:2010sj, Andriot:2016xvq}, it forbids de Sitter solutions for $p=3$, hence the strikethrough entries in Table \ref{tab:before}.

Finally, an extra 10d equation was considered in \cite{Andriot:2016xvq}. Because one has parallel $D_p/O_p$, it is natural to distinguish which directions are parallel and which are transverse to these sources, especially on the internal manifold $\mmm$. In \cite{Andriot:2016xvq} we considered the internal Einstein equation and traced along the parallel directions. This new trace gave important extra constraints: the combination of some curvatures of internal subspaces and $H$-flux components denoted $\mbox{``combi''}=  2 {\cal R}_{||} + 2 {\cal R}_{||}^{\bot} - |H^{(2)}|^2 - 2|H^{(3)}|^2$, and specific components of the spin connection $f^{||}{}_{\bot\bot}$, all related to parallel and transverse directions, and defined in the main text of this paper, have to be non-zero in a de Sitter solution. We include them in Table \ref{tab:before}.

\begin{table}[h]
\begin{center}
\begin{tabular}{|c|c|c|}
    \hline
     & \multicolumn{2}{|c|}{A de Sitter solution requires $T_{10}>0$ and}\\
    \hline
    $p=\dots$ & ${\cal R}_6 \geq 0$  & ${\cal R}_6 <0$ \\
    \hline
    \hline
    3 & \st{$F_1,\, H$} & \st{nothing} \\
    \hline
    4 & $\ \ F_0,\, H$, $\ \ f^{||}{}_{\bot\bot},\, \mbox{combi}\ \ $ & $\ \ F_2$ or $F_0$, $\ \ f^{||}{}_{\bot\bot},\, \mbox{combi}\ \ $ \\
    \hline
    5 &  & $F_1$, $\ \ f^{||}{}_{\bot\bot},\, \mbox{combi}$ \\
    \hline
    6 &  & $F_0$, $\ \ f^{||}{}_{\bot\bot},\, \mbox{combi}$ \\
    \hline
    7 &  &  \\
    \hline
    8 &  &  \\
    \hline
    9 &  &  \\
    \hline
\end{tabular} \caption{Necessary ingredients for the existence of a classical de Sitter solution with parallel $D_p/O_p$, before this work. The entries are defined in the text. An empty box or a strikethrough text mean a complete no-go theorem in the corresponding case.}\label{tab:before}
\end{center}
\end{table}

Studying the second derivatives of the potential has brought further constraints due to the stability requirement: see a summary in \cite{Wrase:2010ew, Shiu:2011zt}. Finally, the case of intersecting sources was also studied, using either approach. For a summary of conditions in 4d with potential, see \cite{Shiu:2011zt}. Further results for intersecting sources were obtained in 10d in \cite{Andriot:2017jhf}, by including the sourced Bianchi identity and the Einstein trace along internal parallel directions: in particular, any combination of intersecting $D_3/O_3$ and $D_7/O_7$ was excluded, and other combinations were constrained.

\subsection*{The present work: ${\cal R}_6 <0$ and the third scalar field}

There is first a simple reasoning to be made to exclude completely the possibility of having ${\cal R}_6 \geq 0$ in Table \ref{tab:before}. For $p=4$, one of the non-sourced flux Bianchi identity (due to the absence of $p=8$ sources) is $\d F_0=0$. Since $F_0$ is a scalar, it has to be constant. However, the $O_4$ projection requires $F_0$ to be odd: this implies that $F_0=0$ for $p=4$ solutions. As one can read in Table \ref{tab:before}, a classical de Sitter solution with parallel sources of single size $p$ is then bound to have ${\cal R}_6 <0$.

\begin{table}[h]
\begin{center}
\begin{tabular}{|c|c|c|}
    \hline
     & \multicolumn{2}{|c|}{A de Sitter solution requires $T_{10}>0$ and}\\
    \hline
    $p=\dots$ & ${\cal R}_6 \geq 0$  & ${\cal R}_6 <0$ \\
    \hline
    \hline
    3 &  &  \\
    \hline
    4 & $\phantom{F_2}$ $\phantom{\ \ f^{||}{}_{\bot\bot},\, f^{\bot}{}_{\bot ||},\, \mbox{combi}}$ & $\ F_2$, $\ \ f^{||}{}_{\bot\bot},\, f^{\bot}{}_{\bot ||},\, \mbox{combi}\ $ \\
    \hline
    5 &  & $F_1$, $\ \ f^{||}{}_{\bot\bot},\, f^{\bot}{}_{\bot ||},\, \mbox{combi}$ \\
    \hline
    6 &  & $F_0$, $\ \ f^{||}{}_{\bot\bot},\, f^{\bot}{}_{\bot ||},\, \mbox{combi}$ \\
    \hline
    7 &  &  \\
    \hline
    8 &  &  \\
    \hline
    9 &  &  \\
    \hline
\end{tabular} \caption{Necessary ingredients for the existence of a classical de Sitter solution with parallel $D_p/O_p$, after this work. The necessity of (specific) $f^{\bot}{}_{\bot ||}$ is shown for group manifolds with constant fluxes.}\label{tab:after}
\end{center}
\end{table}

In \cite{Danielsson:2012et}, it was proposed to introduce a third scalar field $\sigma$, designed to distinguish between internal parallel and transverse directions. As we show in this paper, the extremum condition for this third scalar (combined with the previous equations) turns out to reproduce the extra 10d equation used previously in \cite{Andriot:2016xvq, Andriot:2017jhf} to obtain new constraints,
\beq
\begin{array}{r|c|l}
\del_{\sigma} V|_0=0 & \overset{\eqref{3eq}}{\longleftrightarrow} & \mbox{trace of Einstein eq. along internal parallel directions}
\end{array}
\ , \label{1eq}
\eeq
hence providing a 4d counterpart. The potential dependence on $\sigma$ was sketched for $p=6$ in \cite{Danielsson:2012et} (see \cite{Obied:2018sgi} for partial results for $p=4$), but a rigourous treatment needs the decomposition of all fields along parallel and transverse directions, in particular the curvature terms introduced in \cite{Andriot:2016xvq}. We derive here completely this potential $V(\rho,\tau,\sigma)$: it allows not only to reproduce the 10d results of \cite{Andriot:2016xvq} but also to study the stability in these three fields.

There is an important motivation in studying this stability: it was argued in \cite{Danielsson:2012et} that the tachyon observed in known de Sitter solutions should be a combination of these three fields, meaning that studying their stability should be enough to find a systematic tachyon, i.e.~a general no-go theorem against stable de Sitter solutions. This point was checked successfully on some explicitly given solutions of \cite{Danielsson:2011au} (see also \cite{Junghans:2016uvg}). In the present work, having derived the formalism on a general manifold in Section \ref{sec:gal}, we argue that it is difficult to conclude in full generality. We then restrict in Section \ref{sec:gpnogo} to group manifolds with constant fluxes, the setting in which de Sitter solutions have been found. To our surprise, we are not able to conclude on a full no-go theorem, contrary to the argument of \cite{Danielsson:2012et}. A reason could be that we are dealing with parallel sources, while solutions of \cite{Danielsson:2011au} have intersecting ones. Rather, we obtain stringent constraints on the value of a ratio of structure constants, $\lambda= -\frac{\delta^{cd}   f^{b_{\bot}}{}_{a_{||}  c_{\bot}} f^{a_{||}}{}_{ b_{\bot} d_{\bot}} }{\tfrac{1}{2} \delta^{ab} \delta^{cd} \delta_{i j}  f^{i_{||}}{}_{a_{\bot} c_{\bot}} f^{j_{||}}{}_{b_{\bot} d_{\bot}}}$, for a solution to {\it exist}, leaving only a small region of parameter space where this remains possible, and restricting the set of appropriate group manifolds. In addition, we identify a small island in this region where any de Sitter solution could have all three scalars {\it stabilized}. We comment more on these results, and their implications for the swampland, in Section \ref{sec:ccl}. The new constraints are already summarized in Table \ref{tab:after}.

\section{The potential, its derivatives and the Bianchi identity}\label{sec:gal}

\subsection{From 10d to 4d}

We start with ten-dimensional (10d) type II supergravities with $D_p$-branes and orientifold $O_p$-planes. The $D_p/O_p$ are collectively called sources, since they source RR fluxes. The 10d action is given as follows in 10d string frame
\beq
{\cal S}= \frac{1}{2 \kappa_{10}^2} \int \d^{10} x \sqrt{|g_{10}|} e^{-2\phi} \left( L_{NSNS} + L_{RR} + L_{{\rm sources}} \right) \ , \label{S10d}
\eeq
where we follow the conventions and framework of \cite{Andriot:2016xvq}. We are interested in solutions with a metric of the form
\beq
\d s^2_{10}= \d s^2_4 + \d s^2_6 \ ,\quad \d s^2_4 = e^{2A(y)} \tilde{g}_{\mu\nu} (x) \d x^\mu \d x^\nu \ ,\quad \d s^2_6= g_{mn} (y) \d y^m \d y^n \ ,\label{10dmetric}
\eeq
where $e^A$ is the warp factor. With this compactification ansatz, the 4d metric is that of a maximally symmetric space-time (e.g.~de Sitter) and the 6d one is that of a compact (Euclidian) internal manifold $\mmm$. The sources considered are of one size $p$ with $3 \leq p \leq 8$, and are space-filling in 4d. They thus wrap $p-3$ internal directions and are transverse to $9-p$ other internal directions. In this work, we consider the sources to be parallel to each other, such that all these directions are the same. As detailed in \cite{Andriot:2016xvq}, we further assume that the sets of parallel and transverse internal directions are globally distinct, in the orthonormal or flat basis. This amounts to have the structure group of the tangent bundle of $\mmm$ included in $O(p-3) \times O(9-p)$, which is the case for fiber bundles or group manifolds. This translates into the following definitions
\beq
\d s_6^2 = \delta_{ab} e^a e^b = \d s_{||}^2 + \d s_{\bot}^2 \ ,\quad  \d s_{||}^2 \equiv e^{a_{||}}{}_{m} e^{b_{||}}{}_{n} \delta_{ab} \d y^m \d y^n \ ,\quad \d s_{\bot}^2 \equiv e^{a_{\bot}}{}_{m} e^{b_{\bot}}{}_{n} \delta_{ab} \d y^m \d y^n \ ,
\eeq
with the flat metric $\delta_{ab}$, the one-forms $e^a=e^a{}_m \d y^m$, and the vielbeins $e^{a}{}_{m}$ which depend a priori on all internal coordinates $y$. The two sets of one-forms $\{e^{a_{||}}\}, \{e^{a_{\bot}}\}$ therefore do not mix globally, but can transform among themselves.

Given such a solution or background, we now derive a 4d theory describing the physics of specific fluctuations, appearing in 4d as scalar fields (independent of 6d coordinates) coupled to gravity. The action for those is written in Einstein frame schematically as
\beq
{\cal S}= M_4^2 \int \d^4 x \sqrt{|g_{4E}|} \left( {\cal R}_{4E} - {\rm kin} - \frac{1}{M_4^2} V \right) \ , \label{S4d}
\eeq
with the 4d Planck mass $M_4$, the scalars kinetic terms being ${\rm kin} \sim \del_{\mu} \varphi \del^{\mu} \varphi$, lifting the index with the 4d metric of signature (-,+,+,+), and the scalar potential $V$ giving a mass to $\varphi$ through $\del^2_{\varphi} V$. We study three fluctuations $\rho, \sigma, \tau >0$ with respect to background fields labeled with a ${}^0$ (soon to be dropped). The first two are internal metric fluctuations, also defined through the vielbeins as follows
\beq
\d s_6^2 = \rho \left( \sigma^A (\d s_{||}^2)^0 + \sigma^B (\d s_{\bot}^2)^0 \right) \ , \ e^{a_{||}}{}_{m} = \sqrt{\rho \sigma^A}\, (e^{a_{||}}{}_{m})^0 \ ,\ e^{a_{\bot}}{}_{m} = \sqrt{\rho \sigma^B}\, (e^{a_{\bot}}{}_{m})^0 \label{fluctmetricN} \ ,
\eeq
with real numbers $A,B$, and the third one is built from a dilaton fluctuation
\beq
\phi = \phi^0 + \delta \phi \ ,\quad \tau = e^{- \delta \phi} \rho^{\frac{3}{2}}  \ . \label{fluctdilN}
\eeq
Considering these three fluctuations was proposed in \cite{Danielsson:2012et}, where $\sigma$ was introduced. The definition of $\tau$ is fixed thanks to the relation to Einstein frame as we detail below. To avoid $\sigma$ entering the 6d metric determinant and $\tau$ definition, one should impose the equation
\beq
A(p-3) + B (9-p) = 0 \ . \label{ABeq}
\eeq
Up to a common rescaling (that will always be possible in our equations), the solution is
\bea
A= p-9 \ ,\ B=p-3 \ , \qquad {\rm i.e.}\ A+B=2(p-6) \ ,\ A-B=-6 \ .\label{ABvalues}
\eea
We present a different definition of $\sigma$ in Appendix \ref{ap:sigma}: as shown there, our results are not modified by this other choice.\footnote{In this work we do not determine the kinetic terms explicitly since we do not need them. The kinetic terms for $\rho$ and $\tau$ have been determined in \cite{Hertzberg:2007wc} and they do not mix. The one for $\sigma$ introduced in \cite{Danielsson:2012et} has not been determined, but the three-dimensional field space metric $g_{\varphi_i \varphi_j}$ is still thought to be positive definite. The values picked for $A$ and $B$ could be justified by requiring a non-mixed kinetic term for $\sigma$: the present choice \eqref{ABeq} makes it likely to happen since it avoids the mixing with $\tau$ and makes $\sigma$ disappear from the 6d determinant. In addition, as argued in Appendix \ref{ap:sigma}, the equation $\del_{\sigma} V = 0$ gives in any case the same condition, up to linear combinations with the other equations, so our results are not affected by a change in $\sigma$. The stability study could be affected by a different definition, but the conditions we derive are anyway known to be only necessary.} Finally, the background is recovered by setting these scalar fields to their background value
\beq
\mbox{Background value:}\quad \rho=\sigma = \tau =1 \ . \label{bckgdN}
\eeq

From now on, we consider for computational simplicity no warp factor, i.e. $A=0$, thus drop the tilde notation introduced in \eqref{10dmetric}. In addition, we consider the background dilaton to be constant and given by $e^{\phi^0}=g_s$. As shown in \cite{Andriot:2016xvq}, a standard ansatz relating the dilaton to the warp factor makes all derivatives of $A$ and $\phi$ disappear from the equations of motion; this is effectively the situation we are reproducing here with this simplification. Even though this is also sometimes referred to as a ``smearing approximation'', one may rather consider this absence of $A$ and $\phi$ as considering integrated equations (see Section II of \cite{Andriot:2018wzk} for a discussion of this point). Then, fluctuating the first part of the 10d action given in terms of
\beq
L_{NSNS} = {\cal R}_{10} + 4 |\del \phi|^2 - \frac{1}{2} |H|^2 \ ,\label{LNSNS}
\eeq
where for any $q$-form $A_q$ we denote $|A_q|^2=A_{q\, M_1\dots M_q}A_{q\, N_1\dots N_q}g^{M_1 N_1} \dots g^{M_q N_q} /q!$, we obtain
\beq
\sqrt{|g_{10}|} e^{-2\phi}  ({\cal R}_{10} + 4 |\del \phi|^2) = g_s^{-2} \sqrt{|g_{4}|} \sqrt{|g_{6}^0|} \rho^3 e^{-2\delta\phi} ( {\cal R}_{4} +  {\cal R}_{6} (\rho,\sigma) - {\rm kin} ) \ ,
\eeq
from which we infer the rescaling to Einstein frame, and the definition of the 4d Planck mass
\beq
g_{\mu \nu} = \tau^{-2} g_{4E \, \mu \nu}  \ , \quad M_4^2 = \frac{1}{2\kappa_{10}^2} \int \d y^6 \sqrt{|g_6^0|}\ g_s^{-2} \ .\label{EM}
\eeq
We then determine the potential $V(\rho,\sigma,\tau)$ in \eqref{S4d}, starting from the 10d action \eqref{S10d}: combining the above, we deduce
\beq
V = - \frac{1}{2 \kappa_{10}^2} g_s^{-2} \tau^{-2} \int \d y^{6} \sqrt{|g_{6}^0|} \left( L_{6NSNS} + L_{RR} + L_{{\rm sources}} \right) \ , \label{potfirst}
\eeq
where the three $L$ should be fluctuated with respect to $\rho,\sigma,\tau$, and $L_{6NSNS}$ only contains ${\cal R}_6$ and $H$ contributions.

To compute this potential, we need the compactification ansatz of the (background) 10d fields, presented in \cite{Andriot:2016xvq}. Of particular importance are the following 10d fluxes, written as $F_4^{10} = F_4^4 + F_4$ and $F_5^{10} = F_5^4 + F_5$, with $F_5^4=H_4 \w f_5$, where $F_4^4$ and $H_4$ are 4d four-forms, $F_4$, $F_5$ and $f_5$ are purely 6d forms, $f_5$ being a one-form. A priori, one would allow all components of all fluxes off-shell, and fluctuate all of them, but we anticipate on having a background preserving 4d Lorentz invariance, and therefore consider only $F_4^4$ and $H_4$ as 4d fluxes; all other fluxes are purely internal \cite{Andriot:2016xvq}. $F_4^4$ and $H_4$ (only one of them should be considered, depending on the theory IIA or IIB) are then 4d fields, in addition to the 4d metric and the scalar fields.\footnote{$F_4^4$ and $H_4$ still have the freedom to be proportional to a function of 6d coordinates (which we will include), so are not purely 4d fields strictly speaking. In the following actions, one should then keep the integral over $\mmm$ instead of absorbing it into $M_4^2$. We simplify here the notation in the discussion, not writing out this integral, anticipating this way a notation simplification to be made later.} After fluctuation, the 4d action obtained from the 10d one is
\beq
{\cal S}= M_4^2 \int \d x^{4} \sqrt{|g_{4}|} \left( \tau^2 {\cal R}_4 - \frac{1}{2} \rho^{3} g_s^2 \left( |F_4^4|^2 + \frac{1}{2}|H_4|^2 |f_5(\rho,\sigma)|^2 \right) \right) + \dots \ ,
\eeq
where the dots include contributions from all other (internal) fields, and where for simplicity we do not display the dependence of the internal $f_5$ on $\rho, \sigma$. Note that the squares of the 4d fluxes, defined formally below \eqref{LNSNS}, will actually be negative as expressed in \eqref{VU}, due to the 4d metric signature. We go to Einstein frame, and now denote the squares of the 4d fluxes with an ${}^E$, indicating the use of the Einstein metric. The 4d action is then
\beq
\hspace{-0.7cm} {\cal S}= M_4^2 \int \d x^{4} \sqrt{|g_{4E}|} \left( {\cal R}_{4E} - \frac{1}{2} \tau^4 \rho^{3} g_s^2 \left( |F_4^4|^{2E} + \frac{1}{2}|H_4|^{2E} |f_5(\rho,\sigma)|^2 \right) - {\rm kin} - \frac{1}{M_4^2} U \right) \label{S4dUN}
\eeq
This action, expressed in terms of a different potential $U(\rho,\sigma,\tau)$, is the one needed to derive the 4d Einstein equation as we will do in \eqref{Einstein4d}; we comment more there on this whole procedure. The 4d fluxes can be set to their background value because we do not consider their dynamics: the only fields considered are the 4d metric and the three scalar fields $\rho,\sigma,\tau$. The fluxes background values are fixed as follows due to 4d Lorentz invariance \cite{Andriot:2016xvq}: $F_4^{4\, 0}= {\rm vol}_4 \w *_6 F_6^0$, $H_4^{0}= {\rm vol}_4$, $f_5^0=-*_6 F_5^0$, the background 4d volume form ${\rm vol}_4$ being the same in string and Einstein frame, and the background $f_5^0$ being fixed by the on-shell self-duality constraint of $F_5^{10}$. The dynamics of the three scalars is governed by a potential as defined in \eqref{S4d}, that is then given by
\beq
\frac{1}{M_4^2}\ V =  \frac{1}{M_4^2}\ U - \frac{1}{2} \tau^4 \rho^{3} g_s^2 \left( |F_6^0|^{2} + \frac{1}{2} |(*_6 F_5^0)(\rho,\sigma)|^2 \right) \ .\label{VU}
\eeq

We finally focus on the internal fluxes: any internal form gets decomposed on the basis $\{e^{a_{||}}\}, \{e^{a_{\bot}}\}$ and we denote $F_q^{(n)}$ the component of a flux $F_q$ with $n$ internal parallel flat indices, $0 \leq n \leq p-3$. More explicitly, we have as in \cite{Andriot:2016xvq}
\beq
F_q = \frac{1}{q!} F^{(0)}_{a_{1\bot} \dots a_{q\bot}} e^{a_{1\bot}} \w \dots \w e^{a_{q\bot}} + \frac{1}{(q-1)!} F^{(1)}_{a_{1||} a_{2\bot} \dots a_{q\bot}} e^{a_{1||}} \w e^{a_{2\bot}} \w \dots \w e^{a_{q\bot}} + \dots
\eeq
and $|F_q|^2 = \sum_n |F_q^{(n)}|^2$. Thanks to this decomposition, we extract the dependence on $\sigma$. We now compute from \eqref{potfirst} the contributions from internal fields giving the potential $U$, to eventually obtain with \eqref{VU}
\bea
V = \frac{1}{2\kappa_{10}^2} g_s^{-2} & \int \d y^6 \sqrt{|g_6^0|} \bigg[ - \tau^{-2} \bigg( \rho^{-1} {\cal R}_6(\sigma) -\frac{1}{2} \rho^{-3} \sum_n \sigma^{-An-B(3-n)} |H^{(n)0}|^2 \bigg) \label{pot0N}\\
& \phantom{ \int \d y^6 \sqrt{|g_6^0|} \bigg[ } - g_s \tau^{-3} \rho^{\frac{p-6}{2}} \sigma^{B\frac{p-9}{2}} \frac{T_{10}^0}{p+1} \nn\\
& +\frac{1}{2} g_s^2 \bigg( \tau^{-4} \sum_{q=0}^{4} \rho^{3-q} \sum_n  \sigma^{-An-B(q-n)} |F_q^{(n)0}|^2  - \tau^{4} \rho^3 |F_6^0|^2 \nn\\
&  \phantom{+\frac{1}{2} g_s^2  \bigg(}\ + \frac{1}{2}  \sum_n   (\tau^{-4} \rho^{-2} \sigma^{-An-B(5-n)} |F_5^{(n)0}|^2 - \tau^{4} \rho^2 \sigma^{-An-B(1-n)} |(*_6 F_5)^{(n)0}|^2 ) \bigg) \bigg] \ , \nn
\eea
where only even/odd RR fluxes should be considered in IIA/IIB. The source terms $T_{10}$, defined in \cite{Andriot:2016xvq}, contains the sources charges and the $\delta$ localizing them in their transverse directions. This term only comes from the DBI action; the WZ piece of the source action is topological, i.e.~does not depend on the dilaton nor the metric, so does not contribute here.

We are left with extracting the $\sigma$ dependence of ${\cal R}_6$. One could use the expression of ${\cal R}_6$ in terms of the $f^{a}{}_{bc} = 2 e^a{}_m \del_{[b} e^m{}_{c]}$, deduced from the expression in terms of the Levi-Civita spin connection
\beq
{\cal R}_6= 2 \delta^{ab} \del_a f^c{}_{bc} - \delta^{cd} f^a{}_{ac} f^b{}_{bd} - \frac{1}{4} \left( 2 \delta^{cd} f^a{}_{bc} f^b{}_{ad} + \delta_{ad} \delta^{be} \delta^{cg} f^a{}_{bc} f^d{}_{eg} \right) \ . \label{Rfabc}
\eeq
However, we prefer to build ${\cal R}_6$ from the Ricci tensor, in view of matching with the 10d equations of \cite{Andriot:2016xvq}. We thus consider
\beq
{\cal R}_6 = \delta^{cd} \left({\cal R}_{cd=c_{||} d_{||}} + {\cal R}_{cd=c_{\bot} d_{\bot}}  \right) \ .
\eeq
We obtain
\bea
& 2 \delta^{cd} {\cal R}_{cd=c_{||} d_{||}} = 2 {\cal R}_{||} +  2 \delta^{ab} \del_{\tilde{a}_{\bot}}  f^{c_{\bot}}{}_{c_{\bot}b_{\bot}} + 2 {\cal R}_{||}^{\bot} + |f^{{}_{||}}{}_{{}_{\bot} {}_{\bot}}|^2 \label{Rpartrace}\\
&  2 \delta^{cd} {\cal R}_{cd=c_{\bot} d_{\bot}} = 2 {\cal R}_{\bot} +  2  \delta^{ab} \del_{\tilde{a}_{||}}  f^{c_{||}}{}_{c_{||}b_{||}} + 2 {\cal R}_{\bot}^{||} + |f^{{}_{\bot}}{}_{{}_{||} {}_{||}}|^2  \ ,
\eea
where we used $f^{a}{}_{ab} = 0$ (compactness without boundary of $\mmm$), i.e.~$f^{a_{\bot}}{}_{a_{\bot} b}=- f^{a_{||}}{}_{a_{||} b}$, and we define
\bea
2 {\cal R}_{||}=& 2 R_{||} + 2 \delta^{cd} \left( {\del}_{c_{||}} f^{a_{||}}{}_{a_{||} d_{||}} + f^{a_{||}}{}_{a_{||} c_{||}} f^{b_{||}}{}_{b_{||} d_{||}}  \right)   \\
2 {\cal R}_{||}^{\bot}= & - \delta^{cd} \left( f^{b_{\bot}}{}_{a_{\bot} c_{||}} f^{a_{\bot}}{}_{b_{\bot} d_{||}} + 2 f^{b_{||}}{}_{a_{\bot}  c_{||}} f^{a_{\bot}}{}_{ b_{||} d_{||}} \right) \\
& - \delta^{bg}\delta^{cd}\delta_{ah} \left( f^{h_{\bot}}{}_{g_{\bot} c_{||}} f^{a_{\bot}}{}_{b_{\bot} d_{||}} +  f^{h_{\bot}}{}_{g_{||} c_{||}} f^{a_{\bot}}{}_{b_{||} d_{||}} \right)\nn\\
|f^{{}_{||}}{}_{{}_{\bot} {}_{\bot}}|^2 =& \frac{1}{2} \delta^{bj}\delta^{ah} \delta_{i g}  f^{i_{||}}{}_{a_{\bot} j_{\bot}} f^{g_{||}}{}_{h_{\bot} b_{\bot}} \ , \\
2 {\cal R}_{\bot}=& 2 R_{\bot} + 2 \delta^{cd} \left( {\del}_{c_{\bot}} f^{a_{\bot}}{}_{a_{\bot} d_{\bot}} + f^{a_{\bot}}{}_{a_{\bot} c_{\bot}} f^{b_{\bot}}{}_{b_{\bot} d_{\bot}}  \right) \\
2 {\cal R}_{\bot}^{||}= & - \delta^{cd} \left( f^{b_{||}}{}_{a_{||} c_{\bot}} f^{a_{||}}{}_{b_{||} d_{\bot}} + 2 f^{b_{\bot}}{}_{a_{||}  c_{\bot}} f^{a_{||}}{}_{ b_{\bot} d_{\bot}} \right) \\
& - \delta^{bg}\delta^{cd}\delta_{ah} \left(  f^{h_{||}}{}_{g_{||} c_{\bot}} f^{a_{||}}{}_{b_{||} d_{\bot}} +  f^{h_{||}}{}_{g_{\bot} c_{\bot}} f^{a_{||}}{}_{b_{\bot} d_{\bot}} \right)\nn \\
|f^{{}_{\bot}}{}_{{}_{||} {}_{||}}|^2 =& \frac{1}{2} \delta^{bj}\delta^{ah} \delta_{i g} f^{i_{\bot}}{}_{a_{||} j_{||}} f^{g_{\bot}}{}_{h_{||} b_{||}} \ .
\eea
The quantities ${\cal R}_{||}$ and ${\cal R}_{||}^{\bot}$ already appeared in \cite{Andriot:2016xvq}, where we assumed $f^{c_{\bot}}{}_{c_{\bot}b_{\bot}}=0$, simplifying $\delta^{cd} {\cal R}_{cd=c_{||} d_{||}}$. The quantities $R_{||}$ and $R_{\bot}$ are the Ricci scalars for the pure parallel or transverse subspaces, obtained using the formula \eqref{Rfabc} with only parallel or transverse indices.\footnote{The derivative terms following ${\cal R}_{||}$ and ${\cal R}_{\bot}$ in \eqref{R6final} would each cancel against a term inside the latter, making all derivatives of $f$ cancel; this can be understood because the general formula for the Ricci scalar only involves $\del f^a{}_{ab}$, and the trace vanishes. The squares following ${\cal R}_{||}^{\bot}$ and ${\cal R}_{\bot}^{||}$ could also simplify the latter. Finally, the terms ${\cal R}_{||} - R_{||}$ and ${\cal R}_{\bot} - R_{\bot}$ would cancel some terms in $R_{||}$ and $R_{\bot}$. We nevertheless preserve these notations and quantities to compare to 10d equations of \cite{Andriot:2016xvq}.} We finally obtain
\beq
{\cal R}_6 = {\cal R}_{||} + \delta^{ab} \del_{a_{||}}  f^{c_{||}}{}_{c_{||}b_{||}}  + {\cal R}_{\bot} +  \delta^{ab} \del_{a_{\bot}}  f^{c_{\bot}}{}_{c_{\bot}b_{\bot}} + {\cal R}_{||}^{\bot} + \frac{1}{2} |f^{{}_{\bot}}{}_{{}_{||} {}_{||}}|^2 + {\cal R}_{\bot}^{||} + \frac{1}{2} |f^{{}_{||}}{}_{{}_{\bot} {}_{\bot}}|^2 \ . \label{R6final}
\eeq
We now deduce the dependence on $\sigma$: thanks to the scaling of the vielbeins, one simply needs to look at the indices of the $f^a{}_{bc}$. We obtain
\bea
{\cal R}_6 (\sigma) = &\ \sigma^{-B} \left( {\cal R}_{\bot} +  \delta^{ab} \del_{a_{\bot}}  f^{c_{\bot}}{}_{c_{\bot}b_{\bot}} + {\cal R}_{\bot}^{||} + |f^{{}_{||}}{}_{{}_{\bot} {}_{\bot}}|^2 \right)^0 \label{R6sigmaN}\\
& + \sigma^{-A} \left( {\cal R}_{||} + \delta^{ab} \del_{a_{||}}  f^{c_{||}}{}_{c_{||}b_{||}}  +  {\cal R}_{||}^{\bot} +|f^{{}_{\bot}}{}_{{}_{||} {}_{||}}|^2 \right)^0 \nn\\
& - \frac{1}{2} \sigma^{-2A+B} |f^{{}_{\bot} 0}{}_{{}_{||} {}_{||}}|^2 - \frac{1}{2} \sigma^{-2B+A} |f^{{}_{||} 0}{}_{{}_{\bot} {}_{\bot}}|^2 \nn\\
= &\ \sigma^{-B}\ {\cal R}_6^0 + (\sigma^{-A} - \sigma^{-B}) \left( {\cal R}_{||} + \delta^{ab} \del_{a_{||}}  f^{c_{||}}{}_{c_{||}b_{||}}  +  {\cal R}_{||}^{\bot} +|f^{{}_{\bot}}{}_{{}_{||} {}_{||}}|^2 \right)^0 \label{R6sigmaN2}\\
& - \frac{1}{2} (\sigma^{-2A+B} - \sigma^{-B}) |f^{{}_{\bot} 0}{}_{{}_{||} {}_{||}}|^2 - \frac{1}{2} (\sigma^{-2B+A} - \sigma^{-B}) |f^{{}_{||} 0}{}_{{}_{\bot} {}_{\bot}}|^2 \nn \\
= &\ \sigma^{-B}\ {\cal R}_6^0 + (\sigma^{-A} - \sigma^{-B}) \left( {\cal R}_{||} + \delta^{ab} \del_{a_{||}}  f^{c_{||}}{}_{c_{||}b_{||}}  +  {\cal R}_{||}^{\bot} \right)^0 \label{R6sigmaN3}\\
& - \frac{1}{2} (\sigma^{-2A+B} - 2\sigma^{-A} + \sigma^{-B}) |f^{{}_{\bot} 0}{}_{{}_{||} {}_{||}}|^2 - \frac{1}{2} (\sigma^{-2B+A} - \sigma^{-B}) |f^{{}_{||} 0}{}_{{}_{\bot} {}_{\bot}}|^2 \nn \ .
\eea
${\cal R}_{\bot}^{||} $ and ${\cal R}_{||}^{\bot} $ do not have a uniform scaling, hence the addition and substraction of $|f|^2$ terms.

We now have all ingredients in the potential \eqref{pot0N}. Before studying its variation, let us simplify the notations. First, each background quantity in $V$ is integrated over $\mmm$: we now use the same symbol for the integrated quantity and the local quantity (times the internal volume); if the quantity is constant over the manifold, this replacement is an equality. A second simplification is to drop the background labels ${}^0$. Overall, we simplify notations as follows
\beq
\frac{\int \d y^6 \sqrt{|g_6^0|} |H^{(n)0}|^2 }{\int \d y^6 \sqrt{|g_6^0|}} \rightarrow |H^{(n)0}|^2 \rightarrow |H^{(n)}|^2 \ . \label{notation}
\eeq
We rewrite the potential as follows, introducing the convenient notation $\tV$, with ${\cal R}_6(\sigma)$ given in \eqref{R6sigmaN}
\begin{empheq}[innerbox=\fbox, left=\!\!\!\!\!]{align}
\tV = \frac{1}{M_4^2}\ V = & - \tau^{-2} \bigg( \rho^{-1} {\cal R}_6(\sigma) -\frac{1}{2} \rho^{-3} \sum_n \sigma^{-An-B(3-n)} |H^{(n)}|^2 \bigg) \label{potN}\\
& - g_s \tau^{-3} \rho^{\frac{p-6}{2}} \sigma^{B\frac{p-9}{2}} \frac{T_{10}}{p+1} \nn\\
& \hspace{-0.4in} +\frac{1}{2} g_s^2 \bigg( \tau^{-4} \sum_{q=0}^{4} \rho^{3-q} \sum_n  \sigma^{-An-B(q-n)} |F_q^{(n)}|^2  - \tau^{4} \rho^3 |F_6|^2 \nn\\
& \hspace{-0.4in} \phantom{+\frac{1}{2} g_s^2  \bigg(} + \frac{1}{2}  \sum_n   (\tau^{-4} \rho^{-2} \sigma^{-An-B(5-n)} |F_5^{(n)}|^2 - \tau^{4} \rho^2 \sigma^{-An-B(1-n)} |(*_6 F_5)^{(n)}|^2 ) \bigg) \nn
\end{empheq}

An extra equation will play an important role in our study: a flux Bianchi identity. Given we consider parallel $D_p/O_p$ of one size $p$, there is only one RR flux $F_{k=8-p}$ which is sourced by them, as seen in the corresponding Bianchi identity
\bea
&\d F_k - H \w F_{k-2} = \varepsilon_p\, \frac{T_{10}}{p+1} {\rm vol}_{\bot} \ , \label{BI2}\\
& \mbox{for}\ 0 \leq k=8-p \leq 5\ ,\ \varepsilon_p=(-1)^{p+1} (-1)^{\left[\frac{9-p}{2} \right]} \ ,\nn
\eea
with $F_{-1} =F_{-2}=0$. Projecting this equation on the transverse directions, one rewrites it into
\bea
g_s \frac{T_{10}}{p+1} = & - \frac{1}{2} \left|*_{\bot}H^{(0)} + \varepsilon_p g_s F_{k-2}^{(0)} \right|^2  - \frac{1}{2} \sum_{a_{||}} \left| *_{\bot}( \d e^{a_{||}})|_{\bot} - \varepsilon_p g_s\, \iota_{a_{||}} F_k^{(1)} \right|^2 \label{BI} \\
& + \varepsilon_p g_s (\d F_{k}^{(0)} )_{\bot} + \frac{1}{2}|H^{(0)}|^2 + \frac{1}{2} g_s^2 |F_{k-2}^{(0)}|^2 + \frac{1}{2} g_s^2 |F_{k}^{(1)}|^2 + \frac{1}{2} |f^{{}_{||}}{}_{{}_{\bot} {}_{\bot}}|^2 \ , \nn
\eea
where $(\d e^{a_{||}})|_{\bot} = -\frac{1}{2} f^{a_{||}}{}_{b_{\bot}c_{\bot}} e^{b_{\bot}}\w e^{c_{\bot}}$, the contraction defined as $\iota_a e^b = \delta^b_a$, and we refer to \cite{Andriot:2016xvq} for more detail. This useful equation does not have a 4d interpretation, and is rather viewed as a 10d (stringy) input or consistency requirement.

\subsection{Einstein equation, first and second derivatives}\label{sec:equations}

We now derive the Einstein equation and the derivatives of the potential. We will consider these equations in their background value, given by $\rho=\sigma = \tau =1$ \eqref{bckgdN},\footnote{Setting the scalar fields to these values can be understood equivalently as follows. For a scalar $\varphi$, the quantities $V, \varphi \del_{\varphi} V, \varphi^2 \del^2_{\varphi} V$ have the same powers in $\varphi$: each background quantity $X$ appearing in the latter scales in the same manner $\varphi^q X$. If the equation $\del_{\varphi} V = 0$ to be considered admits a solution $\varphi_0$, one can then redefine the background quantity by absorbing the scalars $\varphi_0^q X \rightarrow X$. This amounts to set the value $\varphi_0=1$ and fix this way the background quantities. So we set $\rho=\sigma = \tau =1$ in the background as argued in \eqref{bckgdN}.} and the equations will then only involve background fields. We first obtain
\bea
\tau \del_\tau \tV|_0 = &\ 2  {\cal R}_6 -  |H|^2 +3 g_s \frac{T_{10}}{p+1} -2 g_s^2 \sum_{q=0}^{6} |F_q|^2 \ ,\\
\rho \del_\rho \tV|_0 = &\  {\cal R}_6 - \frac{3}{2} |H|^2  - g_s \frac{p-6}{2} \frac{T_{10}}{p+1}  +\frac{1}{2} g_s^2 \sum_{q=0}^{6} (3-q) |F_q|^2   \ , \nn
\eea
where we recall the convenient $\tV = V/M_4^2 $ introduced in \eqref{potN}, and the symbol $|_0$ denotes here the background value of $\varphi \del_{\varphi} \tV$. These two expressions have been obtained using $|F_q|^2= \sum_n |F_q^{(n)}|^2$ and $ |F_5|^2 = |*_6 F_5|^2 $.

In the following, we focus on solutions with constant scalar fields, meaning extrema of the potential $V$. To characterize them, the first equation of motion to consider is the Einstein equation. We emphasize that this equation should be derived from \eqref{S4dUN} and not from \eqref{S4d}: it is given as follows for $\rho=\sigma = \tau =1$, dropping the ${}_E$ and ${}^0$ labels
\bea
& {\cal R}_{\mu\nu} - \frac{1}{2} g_s^2 \left(\frac{1}{3!} F^4_{4\, \mu\kappa\lambda\iota} F^{4\ \, \kappa\lambda\iota}_{4\, \nu} + \frac{1}{2} \frac{1}{3!} H_{4\, \mu\kappa\lambda\iota} H^{\ \, \ \kappa\lambda\iota}_{4\, \nu} |f_5|^2 \right) \label{Einstein4d}\\
& -\frac{1}{2} g_{\mu\nu} \left( {\cal R}_{4} - \frac{1}{2} g_s^2 \left( |F_4^4|^{2} + \frac{1}{2}|H_4|^{2} |f_5|^2 \right) - \frac{1}{M_4^2} U|_0 \right) = 0 \nn
\eea
One verifies its trace to be ${\cal R}_4= 2 U|_0 /M_4^2 + g_s^2 (|F_6|^2 + \frac{1}{2}|F_5|^2) $. In the way we defined the potential $V$, it is worth mentioning that one does not have ${\cal R}_4= 2 V|_0 /M_4^2$ here. The difference is due to the two extra 4d fields $F_4^4$ and $H_4$ in our 4d action \eqref{S4dUN}, related to the internal fields $F_6$ and $F_5$.\footnote{The internal fields $F_6$ and $F_5$ rarely appear as non-zero background fluxes in the limit $A=0$ (see e.g.~typical Minkowski solutions in \cite{Andriot:2016ufg}). We believe this is the reason why the difference pointed-out was not noticed previously in the literature studying $V(\rho,\tau)$, as it would have required an explicit solution with these fields non-zero.} The matching of our equations with the 10d equations, shown below, provides a cross-check of our procedure. An alternative procedure, used to derive 4d supergravities, trades these 4d fluxes for scalars, by introducing Lagrange multipliers (see e.g.~appendix E.2 of \cite{Louis:2002ny}), leading to an equivalent though different set of fields and action. As a result, one defines there a different scalar potential, directly related to ${\cal R}_4$ at an extremum, which matches our on-shell expression for ${\cal R}_4$. At the level of equations of motion, the two procedures are thus equivalent. Since we do not aim here at reproducing a full 4d supergravity but only study a few scalar fields, we define $V$ differently, in line with the original derivation of $V(\rho,\tau)$. This allows us below a direct comparison to the 10d equations of motion expressed as well in terms of $F_6$ and $F_5$.

On top of the Einstein equation, we have the first two extrema equations on the potential, giving the following three equations
\bea
({\cal R}_4): \qquad {\cal R}_4= &\ -  2 {\cal R}_6 + |H|^2  - 2 g_s \frac{T_{10}}{p+1}  + g_s^2 \sum_{q=0}^{6} |F_q|^2  \label{eqR4} \\
(\del_\tau V): \qquad 0  = &\ 2  {\cal R}_6 -  |H|^2 +3 g_s \frac{T_{10}}{p+1} -2 g_s^2  \sum_{q=0}^{6} |F_q|^2 \label{eqdeltauV}\\
(\del_\rho V): \qquad 0  = &\  {\cal R}_6 - \frac{3}{2} |H|^2  - g_s \frac{p-6}{2} \frac{T_{10}}{p+1}  +\frac{1}{2} g_s^2 \sum_{q=0}^{6} (3-q) |F_q|^2  \label{eqdelrhoV} \ .
\eea
Interestingly, one verifies that these three equations correspond to the starting three equations of \cite{Andriot:2016xvq}, obtained through 10d equations of motion in string frame. Indeed
\bea
2 \eqref{eqR4} + \eqref{eqdeltauV} &\, \leftrightarrow (2.13) \ {\rm there}\\
-4 \eqref{eqR4} -\tfrac{3}{2} \eqref{eqdeltauV} - \eqref{eqdelrhoV} &\, \leftrightarrow (2.14) \ {\rm there}\\
\eqref{eqR4} &\, \leftrightarrow (2.15) \ {\rm there}
\eea
at least for a constant dilaton and no warp factor.\footnote{Note that there, the equations were local while here, we rather consider the integrated version; we recall our simplifying notations \eqref{notation}.} We can then rederive equations of \cite{Andriot:2016xvq}, such as the following one, here obtained as $\eqref{eqR4} + \eqref{eqdeltauV}$
\beq
{\cal R}_4= g_s \frac{T_{10}}{p+1} - g_s^2 \sum_{q=0}^{6} |F_q|^2 \label{R4T10F} \ .
\eeq
We recall that this equation gives the requirement of having $T_{10}>0$ for de Sitter solutions \cite{Maldacena:2000mw}, satisfied here thanks to the presence of $O_p$.

We turn to the derivative with respect to $\sigma$. We first obtain
\bea
\sigma \del_{\sigma} \tV|_0 = & \ B\ {\cal R}_6 + (A - B) \left( {\cal R}_{||} + \delta^{ab} \del_{a_{||}}  f^{c_{||}}{}_{c_{||}b_{||}}  +  {\cal R}_{||}^{\bot} \right)  + \frac{1}{2} (A-B) |f^{{}_{||}}{}_{{}_{\bot} {}_{\bot}}|^2 \nn\\
& - \frac{1}{2}  \sum_n (An+B(3-n)) |H^{(n)}|^2 - g_s  B\frac{p-9}{2} \frac{T_{10}}{p+1} \label{delVsigmaN}\\
& -\frac{1}{2} g_s^2 \bigg(  \sum_{q=0}^{4}  \sum_n  (An+B(q-n)) |F_q^{(n)}|^2   \nn\\
& \phantom{+\frac{1}{2} g_s^2  \bigg(} + \frac{1}{2}  \sum_n   ( (An+B(5-n)) |F_5^{(n)}|^2 - (An+B(1-n)) |(*_6 F_5)^{(n)}|^2 ) \bigg) \ . \nn
\eea
We rewrite the above as follows
\bea
\sigma \del_{\sigma} \tV|_0 = & \ B \bigg( {\cal R}_6 - \frac{3}{2} |H|^2 - g_s \frac{p-9}{2} \frac{T_{10}}{p+1} -\frac{1}{2} g_s^2 \big(  \sum_{q=0}^{4} q |F_q|^2 + 2 |F_5|^2  \big) \bigg) \\
+ & (A - B) \bigg( {\cal R}_{||} + \delta^{ab} \del_{a_{||}}  f^{c_{||}}{}_{c_{||}b_{||}}  +  {\cal R}_{||}^{\bot} + \frac{1}{2} |f^{{}_{||}}{}_{{}_{\bot} {}_{\bot}}|^2 \nn\\
& \qquad \quad - \frac{1}{2}  \sum_n n |H^{(n)}|^2 -\frac{1}{2} g_s^2 \sum_n  n \big(  \sum_{q=0}^{4}   |F_q^{(n)}|^2 + \frac{1}{2} (  |F_5^{(n)}|^2 - |(*_6 F_5)^{(n)}|^2 ) \big) \bigg) \ . \nn
\eea
Setting it to zero, and using \eqref{eqdelrhoV} and \eqref{R4T10F}, we rewrite it as follows
\bea
(\del_\sigma V): \qquad 0 =&\ \frac{3}{2} B ({\cal R}_4 + g_s^2 |F_5|^2 + 2 g_s^2 |F_6|^2) \label{eqdelsigVN}\\
+ & (A - B) \bigg( {\cal R}_{||} + \delta^{ab} \del_{a_{||}}  f^{c_{||}}{}_{c_{||}b_{||}}  +  {\cal R}_{||}^{\bot} + \frac{1}{2} |f^{{}_{||}}{}_{{}_{\bot} {}_{\bot}}|^2 \nn\\
& \qquad \quad -\frac{1}{2} \sum_n  n \bigg( |H^{(n)}|^2 +  g_s^2 \sum_{q=0}^{4} |F_q^{(n)}|^2 + \frac{g_s^2}{2}  ( |F_5^{(n)}|^2 - |(*_6 F_5)^{(n)}|^2 ) \bigg) \bigg) \ . \nn
\eea
Interestingly, using the values \eqref{ABvalues} for $A, B$, this equation is precisely the same as the trace of the Einstein equation along internal parallel directions, equation (6) in \cite{Andriot:2016xvq}, up to the term $\delta^{ab} \del_{a_{||}}  f^{c_{||}}{}_{c_{||}b_{||}}$ here. While one understands where this difference comes from,\footnote{We see from $\delta^{cd} {\cal R}_{cd=c_{||} d_{||}}$ given in \eqref{Rpartrace} that this term does not arise in that trace, while it is the latter which enters the equation considered in \cite{Andriot:2016xvq}. This term appears in ${\cal R}_6 - \delta^{cd} {\cal R}_{cd=c_{||} d_{||}}$, and arises here because we varied the whole ${\cal R}_6$ with respect to $\sigma$. The difference is then understood because we do not consider exactly the same starting quantity. This term can also be viewed as a mixed contribution between parallel and transverse directions, participating to $\del_\sigma V$, since one has $-\delta^{ab} \del_{a_{||}}  f^{c_{\bot}}{}_{c_{\bot}b_{||}} = \delta^{ab} \del_{a_{||}}  f^{c_{||}}{}_{c_{||}b_{||}}$.} we can assume that this term vanishes
\beq
\delta^{ab} \del_{a_{||}}  f^{c_{||}}{}_{c_{||}b_{||}} = 0 \leftrightarrow \int \d y^6 \sqrt{|g_6^0|} (\delta^{ab} \del_{a_{||}}  f^{c_{||}}{}_{c_{||}b_{||}})^0 = 0 \ , \label{assumption}
\eeq
where the arrow recalls the actual meaning of our notation. This holds at least in the case of group manifolds to be considered in Section \ref{sec:gpnogo}, where $f^{c_{||}}{}_{c_{||}b_{||}}$ is constant. Up to this assumption, we then have reproduced in 4d all 10d equations that have been used in \cite{Andriot:2016xvq}, namely \eqref{eqR4}, \eqref{eqdeltauV}, \eqref{eqdelrhoV}, \eqref{eqdelsigVN}, together with the flux Bianchi identity \eqref{BI}. This shows the equivalence of the two frameworks and allows us to make use of the results and constraints derived in \cite{Andriot:2016xvq}. We will in particular use the requirement, derived there for de Sitter solutions, to have $ |f^{{}_{||}}{}_{{}_{\bot} {}_{\bot}}|^2 \neq 0$.\\

We turn to the second derivatives of the potential. As mentioned below \eqref{S4d}, the square of the mass of a scalar $\varphi$ is given by $\del^2_{\varphi} V$ in our conventions. Therefore to avoid tachyons or flat directions, one requires $\del^2_{\varphi} V > 0 $.\footnote{One may wonder at this stage about an analogue of the Breitenlohner-Freedman (BF) bound that exists for anti-de Sitter space-time, and that allows for slightly negative squares of scalar masses. This bound is derived by analysing the spin 0 unitary representations of the isometry group of the space-time, and the same can be done for de Sitter. The isometry groups are different, but they are both conformal groups. Then one defines in both cases conformal weights $\Delta$, that are subject to (different) constraints. These constraints are obtained by trying to build a well-defined inner product: in the 4d de Sitter case, one gets the restriction that $\Delta \in [0, 3]$ or $\Delta \in \tfrac{3}{2} + \i\, \mathbb{R}$. The Casimir of the algebra defines the square of the mass $m^2$, which is further related to $\Delta$. The constraints then impose that one should have $m^2 \geq 0$ on de Sitter. When $m^2$ grows beyond the opposite of the BF bound, $\Delta$ simply turns to its (allowed) complex values. We refer to \cite{Anninos:2012qw} for more detail. In short, $\del^2_{\varphi} V > 0 $ is the right stability bound to consider on de Sitter.} With several scalar fields, the mass matrix is given in terms of $\del_{\varphi_i} \del_{\varphi_j} V$. With a positive definite field space metric $g_{\varphi_i \varphi_j}$, a (meta)stable solution then amounts to have all eigenvalues of $\del_{\varphi_i} \del_{\varphi_j} V$ positive. As argued in \cite{Andriot:2018wzk}, thanks to Sylvester's criterion (see e.g.~\cite{Shiu:2011zt}), a necessary condition for this positivity is to have the diagonal elements $\del_{\varphi_i}^2 V > 0$. So in this paper, we focus on these diagonal terms. From the potential \eqref{potN}, we obtain
\bea
\hspace{-0.3in} \tau^2 \del^2_\tau \tV|_0 = & - 6 {\cal R}_6 +3 |H|^2 - 12 g_s \frac{T_{10}}{p+1} + g_s^2 \bigg( 10 \sum_{q=0}^{4} |F_q|^2 + 2 |F_5|^2 - 6 |F_6|^2 \bigg) \ , \label{pottautau}\\
\hspace{-0.3in} \rho^2 \del^2_\rho \tV|_0 = & -  2 {\cal R}_6 +6 |H|^2  - g_s  \frac{1}{4} (p-6) (p-8) \frac{T_{10}}{p+1}  \label{potrhorho}\\
& + g_s^2 \bigg(  \sum_{q=0}^{4} \frac{1}{2} (3-q)(2-q) |F_q|^2  + |F_5|^2 - 3 |F_6|^2 \bigg) \ . \nn
\eea
We further compute the derivatives with respect to $\sigma$ and rewrite the result as follows
\bea
\sigma^2 \del^2_{\sigma} \tV|_0 = & \ -\sigma \del_{\sigma} \tV|_0 -B^2\ {\cal R}_6 + (A - B)^2 |f^{{}_{\bot}}{}_{{}_{||} {}_{||}}|^2 \label{deldelVsigmaN}\\
& - (A - B) \bigg( (A+B) ( {\cal R}_{||} + \delta^{ab} \del_{a_{||}}  f^{c_{||}}{}_{c_{||}b_{||}} +  {\cal R}_{||}^{\bot} )  + \frac{1}{2} (3B-A)  |f^{{}_{||}}{}_{{}_{\bot} {}_{\bot}}|^2  \bigg) \nn\\
& + \frac{1}{2}  \sum_n ((A-B)n + 3B)^2 |H^{(n)}|^2 - g_s  B^2\frac{(p-9)^2}{4} \frac{T_{10}}{p+1} \nn\\
& +\frac{1}{2} g_s^2 \bigg(  \sum_{q=0}^{4}  \sum_n  ((A-B)n + qB)^2 |F_q^{(n)}|^2   \nn\\
& \phantom{+\frac{1}{2} g_s^2  \bigg(} + \frac{1}{2}  \sum_n   ( ((A-B)n + 5B)^2 |F_5^{(n)}|^2 - ((A-B)n + B)^2 |(*_6 F_5)^{(n)}|^2 ) \bigg) \ . \nn
\eea

We have now derived all equations to analyse the possibility of getting metastable de Sitter solutions or finding a systematic tachyon, namely the Einstein equation \eqref{eqR4}, the first derivatives \eqref{eqdeltauV}, \eqref{eqdelrhoV}, \eqref{delVsigmaN}, the second derivatives \eqref{pottautau}, \eqref{potrhorho}, \eqref{deldelVsigmaN}, and the Bianchi identity \eqref{BI}. Let us first comment on the situation with only $\rho, \tau$. Extensive studies in the $(\tau,\rho)$-plane, ending with \cite{Shiu:2011zt}, have not been conclusive on the general sign of the second derivatives, in particular regarding the presence of a general tachyon. A way to see this indetermination goes as follows. Using the various equations, we are able to rewrite the second derivatives in terms of only the contributions $g_s^2 |F_q|^2 $ and $ g_s \frac{T_{10}}{p+1}$. We get schematically
\beq
\del^2_\varphi V \sim a\  g_s \frac{T_{10}}{p+1} - b\ g_s^2 |F_q|^2 \ ,\ {\rm with} \ a< b\ .
\eeq
This should be compared with equation \eqref{R4T10F} that measures the difference between these two quantities in terms of the cosmological constant: we see that the configuration $a< b$ is precisely the one that does not allow to conclude.

Considering a third scalar $\sigma$ could then bring further interesting constraints, and maybe indicate a systematic tachyon as argued in \cite{Danielsson:2012et}. However, doing so may as well introduce new unknown quantities which eventually leaves the problem as complicated and inconclusive as before.\footnote{The quantity $|f^{{}_{\bot}}{}_{{}_{||} {}_{||}}|^2$ appears here, especially in the second derivative. There are several arguments to set this quantity to zero: first, $f^{{}_{\bot}}{}_{{}_{||} {}_{||}}$ is morally T-dual to $H^{(3)}$ that should be set to zero to avoid the Freed-Witten anomaly (see \cite{Andriot:2016xvq}). Secondly, on a group manifold, the structure constant $f^{{}_{\bot}}{}_{{}_{||} {}_{||}}$ would be set to zero by an orientifold projection. So our discussion does not consider that quantity.} This is to some extent what we observe here: two new quantities appear, $ {\cal R}_{||} +  {\cal R}_{||}^{\bot} $ and $|f^{{}_{||}}{}_{{}_{\bot} {}_{\bot}}|^2$, in such a way that one is not able to conclude in general. The former should tend to be negative for de Sitter solutions \cite{Andriot:2016xvq}, and the latter non-zero. But they enter with the wrong signs in $\del^2_\sigma V$ to conclude on a tachyon. Schematically, one has for $p\leq 6$
\bea
& \del_\sigma V \sim  -({\cal R}_{||} +{\cal R}_{||}^{\bot}) - \tfrac{1}{2} |f^{{}_{||}}{}_{{}_{\bot} {}_{\bot}}|^2  \label{configsign2}\\
& \del_\sigma^2 V \sim - ( {\cal R}_{||} +  {\cal R}_{||}^{\bot})  + \tfrac{1}{2} |f^{{}_{||}}{}_{{}_{\bot} {}_{\bot}}|^2 \ ,\nn
\eea
where we see that using the first derivative equation does not help. Another equation where the quantity $|f^{{}_{||}}{}_{{}_{\bot} {}_{\bot}}|^2$ appears is the projected Bianchi identity \eqref{BI}, going schematically as
\beq
g_s \frac{T_{10}}{p+1} \sim \tfrac{1}{2} |f^{{}_{||}}{}_{{}_{\bot} {}_{\bot}}|^2 - |{\rm BPS}|^2 + ({\rm flux})^2 \ . \label{BIschematic}
\eeq
Using this relation would introduce a new unknown (or irreplaceable) quantity, the squares of BPS-like relations, with again the wrong sign to conclude on a tachyon. The last place where the two quantities $ {\cal R}_{||} +  {\cal R}_{||}^{\bot} $ and $|f^{{}_{||}}{}_{{}_{\bot} {}_{\bot}}|^2$ appear is within ${\cal R}_6$. The Ricci scalar however contains other terms that cannot be replaced and of indefinite sign. Therefore, without a further assumption, as a geometric simplifying ${\cal R}_6$ or one of the above quantity, it seems we cannot conclude in general. We then turn to the particular case of group manifolds with constant fluxes, which remains a prime setting to find (stable) de Sitter solutions.

\section{Group manifolds, no-go theorems and the stability island}\label{sec:gpnogo}

\subsection{Group manifold setting}

All known de Sitter solutions have been obtained on group manifolds. In addition, it was observed \cite{Danielsson:2012et} that a tachyon is present among the three scalar fields $\tau, \rho, \sigma$ for the solutions of \cite{Danielsson:2011au}, as far as it could be checked. It is thus reasonable to focus on this setting to hopefully reach a general conclusive statement. The first simplification is that the $f^{a}{}_{bc}$ are here constant. Thanks to this, further simplifications occur due to the orientifold projection as detailed in \cite{Andriot:2017jhf}: structure constants with odd numbers of $\bot$ indices vanish. Thanks to \eqref{R6final} and related formulas, ${\cal R}_6$ boils down to
\beq
{\cal R}_6 = {\cal R}_{||} + {\cal R}_{||}^{\bot} - \frac{1}{2} |f^{{}_{||}}{}_{{}_{\bot} {}_{\bot}}|^2  - \delta^{cd}   f^{b_{\bot}}{}_{a_{||}  c_{\bot}} f^{a_{||}}{}_{ b_{\bot} d_{\bot}}  \ . \label{R6final2}
\eeq
More simplifications occur when assuming only constant fluxes. Having a group manifold strongly suggests, because of the Einstein equations, that the flux components (with flat indices) have to be constant (at least with $A=0$ and constant dilaton), even though one should verify in detail whether cancelations between non-constant components could happen. We assume these components are constant from now on: combining this with the orientifold projection has important consequences. While the $H$-flux should be odd under the orientifold involution $\sigma$, it varies according to the flux and source for RR fluxes. It is given by $\sigma(\sum_q F_q) = \pm \alpha (\sum_q F_q)$, where $+$ is for $p=3,6,7$ and $-$ for $p=4,5,8$, and $\alpha(F_q)= (-1)^{\frac{q(q-1)}{2}} F_q$. One deduces the parity of each flux. With the assumption of constant flux (component), and using as well the dimensionality of parallel and transverse spaces to the source, one concludes on which component can be non-zero: the only non-zero (constant) components for each source are
\bea
& O_6: \quad F_0^{(0)}, F_2^{(1)}, F_4^{(2)}, F_6^{(3)}  \Rightarrow (A-B) n + q B = 0 \ \forall q \label{combip=6}\\
& O_5: \quad F_1^{(0)}, F_3^{(1)}, F_5^{(2)} \phantom{, F_6^{(3)}} \Rightarrow (A-B) n + q B = 3-q\ \forall q \\
& O_4: \quad F_2^{(0)}, F_4^{(1)} \phantom{, F_4^{(2)}, F_6^{(3)}} \Rightarrow (A-B) n + q B = 2(3-q)\ \forall q
\eea
where we used the values \eqref{ABvalues} for $A,B$ to compute the combinations on the right. With the notation \eqref{BI2} where $F_{k=8-p}$ is the flux sourced by the $D_p/O_p$, the only non-zero flux components are
\beq
F_{k-2}^{(0)},\ F_k^{(1)},\ F_{k+2}^{(2)},\ F_{k+4}^{(3)} \ , \label{Fkns}
\eeq
with $F_{q\geq 7}=0$ and $F_q^{(n> p-3 )}=0$. The above combinations on the right give the contributions of fluxes to the first and second derivatives of the potential with respect to $\sigma$: in particular for $p=6$, we read from \eqref{combip=6} that the RR fluxes do not contribute to \eqref{delVsigmaN} and \eqref{deldelVsigmaN}, as already noticed in \cite{Danielsson:2012et}. For $p=5$, we are also interested in the $F_5$ contributions. Since there $F_5=F_5^{(2)}$, $*_6 F_5$ is necessarily transverse to the source, i.e.~$*_6 F_5 = (*_6 F_5)^{(0)}$. With the values \eqref{ABvalues} for $A,B$, we deduce
\bea
&\sum_n   ( ((A-B)n + 5B) |F_5^{(n)}|^2 - ((A-B)n + B) |(*_6 F_5)^{(n)}|^2 ) = -4 |F_5|^2 \ ,\\
&\sum_n   ( ((A-B)n + 5B)^2 |F_5^{(n)}|^2 - ((A-B)n + B)^2 |(*_6 F_5)^{(n)}|^2 ) = 0 \ .\nn
\eea
Finally, the $H$-flux only admits $H^{(0)}, H^{(2)}$ for $p=5,6$, and only $H^{(0)}$ for $p=4$. Using this information, in particular \eqref{Fkns}, the Bianchi identity projection \eqref{BI} becomes
\bea
g_s \frac{T_{10}}{p+1} = & - \frac{1}{2} \left|*_{\bot}H^{(0)} + \varepsilon_p g_s F_{k-2}^{(0)} \right|^2  - \frac{1}{2} \sum_{a_{||}} \left| *_{\bot}( \d e^{a_{||}})|_{\bot} - \varepsilon_p g_s\, \iota_{a_{||}} F_k^{(1)} \right|^2 \label{BIsimp} \\
&  + \frac{1}{2}|H^{(0)}|^2 + \frac{1}{2} g_s^2 |F_{k-2}^{(0)}|^2 + \frac{1}{2} g_s^2 |F_{k}^{(1)}|^2 + \frac{1}{2} |f^{{}_{||}}{}_{{}_{\bot} {}_{\bot}}|^2 \ . \nn
\eea
Thanks to these assumptions of group manifolds and constant fluxes, realised in the known examples of solutions, the different expressions simplify and we may reach more conclusive statements.

\subsection{The general linear combination and no-go theorems}\label{sec:nogo}

Even in the previous restrictive setting, it remains difficult to show that one specific scalar field is always tachyonic. What may rather be shown is that the combination of requirements, i.e.~having all three not tachyonic together with a de Sitter extremum, is too strong. To reach such a result, we take a systematic approach and consider a general linear combination of all equations that appeared so far, namely \eqref{eqR4}, \eqref{eqdeltauV}, \eqref{eqdelrhoV}, \eqref{delVsigmaN}, \eqref{pottautau}, \eqref{potrhorho}, \eqref{deldelVsigmaN}, together with the flux Bianchi identity \eqref{BIsimp}. We build this way a combination that leads to an interesting, i.e.~conclusive, condition. With the real coefficients $a, b_{\tau}, b_{\rho}, b_{\sigma}, c_{\tau}, c_{\rho}, c_{\sigma}, d$, we consider
\bea
& a {\cal R}_4 + b_{\tau} \ \tau \del_{\tau} \tV|_0  + b_{\rho} \ \rho \del_{\rho} \tV|_0  + b_{\sigma} \ \sigma \del_{\sigma} \tV|_0 \label{galmethod}\\
& + c_{\tau} \ \tau^2 \del^2_{\tau} \tV|_0 + c_{\rho} \ \rho^2 \del^2_{\rho} \tV|_0 + c_{\sigma} \ \left( \sigma^2 \del^2_{\sigma} \tV|_0 + \sigma \del_{\sigma} \tV|_0 \right)\nn\\
& +\frac{d}{2}  \left|*_{\bot}H^{(0)} + \varepsilon_p g_s F_{k-2}^{(0)} \right|^2  + \frac{d}{2} \sum_{a_{||}} \left| *_{\bot}( \d e^{a_{||}})|_{\bot} - \varepsilon_p g_s\, \iota_{a_{||}} F_k^{(1)} \right|^2 \nn\\
= & - ({\cal R}_{||} + {\cal R}_{||}^{\bot})\ (2a-2b_{\tau} -b_{\rho} -A b_{\sigma} + 6 c_{\tau} + 2 c_{\rho} + A^2 c_{\sigma}) \nn\\
& + \delta^{cd}   f^{b_{\bot}}{}_{a_{||}  c_{\bot}} f^{a_{||}}{}_{ b_{\bot} d_{\bot}}\ (2a-2b_{\tau} -b_{\rho} -B b_{\sigma} + 6 c_{\tau} + 2 c_{\rho} + B^2 c_{\sigma}) \nn\\
& + \frac{1}{2} |f^{{}_{||}}{}_{{}_{\bot} {}_{\bot}}|^2\  (2a-2b_{\tau} -b_{\rho} +(A-2B) b_{\sigma} + 6 c_{\tau} + 2 c_{\rho} + (B^2 +(B-A)(3B-A) ) c_{\sigma} + d ) \nn\\
& +\frac{1}{2} |H^{(0)}|^2 \ (2a-2b_{\tau} -3b_{\rho} -3B b_{\sigma} + 6 c_{\tau} + 12 c_{\rho} + 9B^2 c_{\sigma} +d) \nn\\
& +\frac{1}{2} |H^{(2)}|^2 \ (2a-2b_{\tau} -3b_{\rho} -(2A+B) b_{\sigma} + 6 c_{\tau} + 12 c_{\rho} + (2A+B)^2 c_{\sigma}) \nn\\
& + g_s \frac{T_{10}}{p+1} \frac{1}{4}\ (-8a +12b_{\tau} - 2(p-6) b_{\rho} - 2B (p-9) b_{\sigma} -48 c_{\tau} - (p-6) (p-8) c_{\rho} -B^2 (p-9)^2 c_{\sigma}-4d) \nn\\
& + g_s^2 \sum_{q=0}^4 \frac{1}{2} |F_q|^2\ (2a-4b_{\tau} + (3-q) b_{\rho}  + 20 c_{\tau} + (3-q)(2-q) c_{\rho})  + g_s^2 \frac{d}{2} ( |F_{k-2}^{(0)}|^2 + |F_{k}^{(1)}|^2 ) \nn\\
&- b_{\sigma}  g_s^2 \sum_{q=0}^{4}  \sum_n  \frac{1}{2}|F_q^{(n)}|^2 ((A-B)n + qB)   + c_{\sigma} g_s^2  \sum_{q=0}^{4}  \sum_n \frac{1}{2}|F_q^{(n)}|^2  ((A-B)n + qB)^2  \nn\\
& + g_s^2 \frac{1}{2} |F_5|^2\ (2a-4b_{\tau}  -2 b_{\rho}  + 4 c_{\tau} + 2 c_{\rho}) \nn\\
& - b_{\sigma}\frac{1}{4} g_s^2   \sum_n   ( ((A-B)n + 5B) |F_5^{(n)}|^2 - ((A-B)n + B) |(*_6 F_5)^{(n)}|^2 ) \nn\\
& + c_{\sigma}\frac{1}{4} g_s^2   \sum_n   ( ((A-B)n + 5B)^2 |F_5^{(n)}|^2 - ((A-B)n + B)^2 |(*_6 F_5)^{(n)}|^2 )    \nn\\
& + g_s^2 \frac{1}{2} |F_6|^2\ (2a-4b_{\tau}  -3 b_{\rho}  -12 c_{\tau} -6 c_{\rho}) \nn
\eea
This linear combination is particularly interesting in the case where the right-hand side is made of positive quantities times linear combinations of the coefficients. This is for instance the case when one has ${\cal R}_{||} + {\cal R}_{||}^{\bot} \leq 0$ and $ \delta^{cd}   f^{b_{\bot}}{}_{a_{||}  c_{\bot}} f^{a_{||}}{}_{ b_{\bot} d_{\bot}} =0$, which holds for a nilmanifold in the ``standard basis'' (see below and \cite{Andriot:2017jhf}). Then, if one finds values of the coefficients, verifying $a, c_{\tau}, c_{\rho}, c_{\sigma}, d \geq 0$, and such that the combinations of coefficients on the right-hand side are negative, one would conclude on the absence of a stable de Sitter solution. We make further comments on this method in Appendix \ref{ap:ex}.

We find a first solution to this inequalities problem in the case where
\beq
\delta^{cd}   f^{b_{\bot}}{}_{a_{||}  c_{\bot}} f^{a_{||}}{}_{ b_{\bot} d_{\bot}} \geq 0 \ , \label{ffpos}
\eeq
with the following coefficients
\beq
b_{\tau}= \frac{3}{2}a\ ,\ b_{\rho}= \frac{A+B}{A-B} a\ ,\ b_{\sigma}= \frac{2}{B-A} a \ ,\ d = 4a \ ,\  c_{\tau}= c_{\rho} = c_{\sigma} =0 \ .\label{nilcoef}
\eeq
In other words, we show for $p=4,5,6$ and the values \eqref{ABvalues} of $A,B$ that
\bea
& {\cal R}_4 + \frac{3}{2} \ \tau \del_{\tau} \tV|_0  + \frac{A+B}{A-B} \ \rho \del_{\rho} \tV|_0  + \frac{2}{B-A} \ \sigma \del_{\sigma} \tV|_0  \label{lincombi1}\\
& +2  \left|*_{\bot}H^{(0)} + \varepsilon_p g_s F_{k-2}^{(0)} \right|^2  + 2 \sum_{a_{||}} \left| *_{\bot}( \d e^{a_{||}})|_{\bot} - \varepsilon_p g_s\, \iota_{a_{||}} F_k^{(1)} \right|^2 \nn\\
= & -2 g_s^2 \left( |F_{k+2}|^2 + |F_{k+4}|^2 \right) - 2\, \delta^{cd}   f^{b_{\bot}}{}_{a_{||}  c_{\bot}} f^{a_{||}}{}_{ b_{\bot} d_{\bot}} \ .\nn
\eea
We deduce the following, for classical de Sitter solutions with parallel sources on a group manifold with constant fluxes
\beq
\boxed{\mbox{No-go:}}\quad \mbox{There is no de Sitter solution when}\ \delta^{cd}   f^{b_{\bot}}{}_{a_{||}  c_{\bot}} f^{a_{||}}{}_{ b_{\bot} d_{\bot}} \geq 0 \ . \label{nogo1}
\eeq

We now consider the case where $ \delta^{cd}   f^{b_{\bot}}{}_{a_{||}  c_{\bot}} f^{a_{||}}{}_{ b_{\bot} d_{\bot}} < 0$. We recall that $|f^{{}_{||}}{}_{{}_{\bot} {}_{\bot}}|^2 >0$ is required for de Sitter solutions \cite{Andriot:2016xvq}, so we introduce the parameter $\lambda >0$ such that
\beq
\delta^{cd}   f^{b_{\bot}}{}_{a_{||}  c_{\bot}} f^{a_{||}}{}_{ b_{\bot} d_{\bot}} = - \lambda |f^{{}_{||}}{}_{{}_{\bot} {}_{\bot}}|^2 <0\ . \label{deflambda}
\eeq
In that case, one obtains
\beq
{\cal R}_6 = {\cal R}_{||} + {\cal R}_{||}^{\bot} + \left( \lambda - \frac{1}{2} \right) |f^{{}_{||}}{}_{{}_{\bot} {}_{\bot}}|^2  \ .
\eeq
As shown in the Introduction, a classical de Sitter solution with parallel sources requires ${\cal R}_6 <0$. In addition, the condition \eqref{eqdelsigVN}, obtained here in the 4d approach, and in the 10d work \cite{Andriot:2016xvq}, implies for $p=4,5,6$ on a group manifold that a de Sitter solution must have
\beq
{\cal R}_{||} + {\cal R}_{||}^{\bot} + \frac{1}{2} |f^{{}_{||}}{}_{{}_{\bot} {}_{\bot}}|^2  > 0 \ .\label{conddsigmaV}
\eeq
Combining both conditions gives the requirement, for a classical de Sitter solution with parallel sources on a group manifold with constant fluxes
\beq
\mbox{A de Sitter solution must have $0 < \lambda < 1$.} \label{lambda1}
\eeq
In other words, we get the following no-go theorem
\beq
\boxed{\mbox{No-go:}}\quad \mbox{There is no de Sitter solution when}\ \lambda= -\frac{\delta^{cd}   f^{b_{\bot}}{}_{a_{||}  c_{\bot}} f^{a_{||}}{}_{ b_{\bot} d_{\bot}} }{|f^{{}_{||}}{}_{{}_{\bot} {}_{\bot}}|^2} \geq 1 \ . \label{nogo2}
\eeq
This reasoning, concluding on the requirement \eqref{lambda1}, only involves equations present in our general combination \eqref{galmethod}, so we should be able to reproduce it with some coefficients.\footnote{Note that away from the nilmanifold case, the sign of ${\cal R}_{||} + {\cal R}_{||}^{\bot} $ is not always definite; for that reason, the combination method \eqref{galmethod} could be problematic, unless the linear combination coming with ${\cal R}_{||} + {\cal R}_{||}^{\bot} $, namely $2a-2b_{\tau} -b_{\rho} -A b_{\sigma} + 6 c_{\tau} + 2 c_{\rho} + A^2 c_{\sigma}$, vanishes. It is actually the case for \eqref{lincombi1} and here for \eqref{lincombi2}.} We do so with the following coefficients
\beq
b_{\tau} = \frac{A-5B}{A-3B} \frac{a}{2} \ ,\ b_{\rho}= \frac{a}{3} \ ,\ b_{\sigma} = \frac{2}{3(A-3B)}a \ ,\  c_{\tau}= c_{\rho} = c_{\sigma} = d=0 \ .
\eeq
In other words, we show for $p=4,5,6$, using the values \eqref{ABvalues} for $A,B$
\bea
& {\cal R}_4 + \frac{1}{2} \frac{A-5B}{A-3B} \ \tau \del_{\tau} \tV|_0  + \frac{1}{3} \ \rho \del_{\rho} \tV|_0  + \frac{2}{3(A-3B)} \ \sigma \del_{\sigma} \tV|_0 \label{lincombi2}\\
= & -\frac{2}{p} \left( |f^{{}_{||}}{}_{{}_{\bot} {}_{\bot}}|^2\ (\lambda - 1)  + |H^{(2)}|^2 \right)  - \frac{2}{p}\, g_s^2 \frac{1}{2} \sum_{q=6-p}^p (q - (6- p))\ |F_q|^2   \nn \ ,
\eea
where we used the knowledge on the various flux components allowed by the orientifold. For $\lambda \geq 1$, the right-hand side is clearly negative, so there is no de Sitter solution, as obtained previously.\footnote{Note that the reasoning leading to \eqref{lambda1} could be refined using \eqref{eqdelsigVN} towards the requirement, for de Sitter solutions, $|f^{{}_{||}}{}_{{}_{\bot} {}_{\bot}}|^2\ (\lambda - 1)  + |H^{(2)}|^2 < 0 $, which is consistent with what we obtain here in \eqref{lincombi2}. One may even include RR fluxes in the requirement.}

As a side comment, we note that
\beq
\hspace{-0.1in} f^a{}_{bc}= -f^b{}_{ac} \Rightarrow f^{b_{\bot}}{}_{a_{||}  c_{\bot}} = - f^{a_{||}}{}_{ b_{\bot} d_{\bot}} \Rightarrow \delta^{cd}   f^{b_{\bot}}{}_{a_{||}  c_{\bot}} f^{a_{||}}{}_{ b_{\bot} d_{\bot}} = - 2 |f^{{}_{||}}{}_{{}_{\bot} {}_{\bot}}|^2 \ ,\ {\rm i.e.}\ \lambda=2 \ . \label{l=2}
\eeq
In that case, there is therefore no de Sitter solution. For intersecting sources, de Sitter solutions are however known on manifolds with $f^a{}_{bc}= -f^b{}_{ac}$. An explanation is that the condition \eqref{eqdelsigVN}, used here to conclude negatively for parallel sources, gets modified by an additional source term for intersecting sources \cite{Andriot:2017jhf}, precisely resulting in making \eqref{conddsigmaV} milder, giving more room for de Sitter solutions.

Eventually, we are only left with the case $0 < \lambda < 1$, for which we are not able to conclude. We provide in Appendix \ref{ap:ex} an example of such a situation. Surprisingly, all conclusive statements or constraints we were able to derive did not involve the second derivatives, and are thus all only about the existence of solutions. This goes against the initial intuition that the tachyon should be present among the three scalar fields. This situation could change when turning to intersecting sources instead of parallel ones. We now comment on the restrictions deduced from our results on the underlying Lie algebras and corresponding manifolds.

\subsection*{Consequences on algebras}

The constraints just derived have singled out the quantity $ \delta^{cd}   f^{b_{\bot}}{}_{a_{||}  c_{\bot}} f^{a_{||}}{}_{ b_{\bot} d_{\bot}} $ that we rewrite here in terms of differential forms
\beq
\delta^{cd}   f^{b_{\bot}}{}_{a_{||}  c_{\bot}} f^{a_{||}}{}_{ b_{\bot} d_{\bot}} \, {\rm vol}_{\bot} = \iota_{b_{\bot}} \d e^{a_{||}} \w *_{\bot}  \iota_{a_{||}} \d e^{b_{\bot}} \ ,\label{quantityff}
\eeq
with the contraction $\iota_a e^b = \delta^b_a$. According to the value of this quantity, we obtained no-go theorems on de Sitter solutions. One could deduce from the latter constraints on the choice of group manifold or corresponding algebra, but this requires to fix the basis in which the algebra is given. Indeed, the structure constants are here expressed in a basis suited to describe the directions parallel and transverse to the sources, which may not be a standard basis to express the algebra. Let us then clarify the constraints to be inferred on the algebras by first discussing the various basis.

We refer to ``the standard basis'' as being the one where the structure constants verify $f^a{}_{ab}=0$, without sum on $a$. More precisely, we consider this basis to be the one where semi-simple algebras have fully antisymmetric structure constants, and solvable algebras are given as in \cite{Grana:2006kf, Andriot:2010ju}. In particular, nilpotent algebras in this basis are such that if for $a,b,c$ $f^a{}_{bc} \neq 0$, then $f^b{}_{ac}=0$. Non-nilpotent solvmanifolds in this basis can be understood as having typically two different kinds of fibrations, on top the nilmanifold ones (see e.g.~almost abelian solvmanifolds \cite{Andriot:2010ju}): the fibration is encoded either through hyperbolic rotations, or standard rotations, i.e.~thanks to (hyperbolic) sines and cosines. For the former, one has, for some $a,b,c$, $f^a{}_{bc}= f^b{}_{ac}$ while for the latter, one has $f^a{}_{bc}= -f^b{}_{ac}$, up to radii. Note that these structure constants are not fully antisymmetric.

In this ``standard basis'', it is easy to evaluate the quantity \eqref{quantityff}. For nilmanifolds, it vanishes. For almost abelian solvmanifolds with hyperbolic rotation, one has $f^{b_{\bot}}{}_{a_{||}  c_{\bot}} = f^{a_{||}}{}_{ b_{\bot} d_{\bot}}$, making the quantity \eqref{quantityff} positive. For semi-simple group manifolds as well as almost abelian solvmanifolds with standard rotation, one has $f^a{}_{bc}= -f^b{}_{ac}$ (up to radii), giving $\lambda= 2$ as shown in \eqref{l=2}. Then, {\it all these cases do not allow for de Sitter solutions.}

Unless rescaling structure constants with radii changes these results (for $\lambda >0$), the constraints obtained point towards solvmanifolds with standard rotation fibrations (i.e.~with the good sign for $\lambda$) that are not almost abelian, meaning that they have an additional a nilpotent fibration. We provide in Appendix \ref{ap:ex} an example of this preferred type of group manifold. Further more involved combinations of algebras could be possible, such as the product of a semi-simple one with a nilpotent one.

Another possibility is to have a different basis for the $\{e^{a_{||}}\}$ and $\{e^{a_{\bot}}\}$ than this standard basis of the Lie algebras. The freedom in choosing a basis is captured by $O(6)$, that preserves the metric and changes the Maurer--Cartan forms. However, the quantity \eqref{quantityff} is a tensor, invariant under the transformations $O(p-3) \times O(9-p)$ that preserve the split into $\{e^{a_{||}}\}$ and $\{e^{a_{\bot}}\}$. To get a different value for this quantity, one should then pick a basis related by a different transformation in $O(6)$ to the standard basis. However, a difficulty in using such a non-standard basis of the algebra, having in particular $f^a{}_{ab} \neq 0$ (without sum on $a$), is to make it compatible with the orientifold projection. There is thus not much freedom left in using a non-standard basis for a group manifold, if that one is excluded in the standard basis.

\subsection{The stability island}\label{sec:stab}

In Section \ref{sec:nogo}, we have excluded de Sitter solutions outside a parameter region $0 < \lambda < 1$, defined in \eqref{deflambda}. In this remaining region, we were not able to conclude on the existence of solutions, or even to find tachyons. Here, we go further in the opposite direction by identifying a subregion or ``island'' where all three scalars can be stabilized. What is meant is that we prove for the three scalars that $\del_{\varphi_i}^2 V|_0 > 0$ for any de Sitter solution in this island, where we recall that this is only a necessary condition for stability (see above \eqref{pottautau}).

To achieve this, we use again the general linear combination \eqref{galmethod} but reverse the logic: we look, for one of the three scalars $\varphi_i$, for a set of coefficients $a> 0 \ ,\ c_{\varphi_i} <0 \ ,\ c_{\varphi_{i\neq j}} = 0 \ ,\ d\geq 0$, such that the right-hand side of the combination is always negative. If such a set of coefficients exists, we deduce that $\del_{\varphi_i}^2 V|_0 > 0$ for any de Sitter solution. We first find a realization of this for $\rho$: indeed, one shows, for $p=4,5,6$, recalling $F_{q\geq 7}=0$ and $F_q^{(n> p-3 )}=0$, that
\bea
& \frac{(p-2)^2-3}{2} {\cal R}_4 + \frac{p^2- 6 p +8}{2} \ \tau \del_{\tau} \tV|_0  + (2 p-9) \ \rho \del_{\rho} \tV|_0  -  \ \rho^2 \del^2_{\rho} \tV|_0  \\
= & -2 (p-2) ( |H^{(0)}|^2  + |H^{(2)}|^2 )  - \frac{(p-4)^2}{4} g_s \frac{T_{10}}{p+1}  - \frac{p+3-(p-7)^2}{2} g_s^2 |F_{10-p}^{(2)}|^2 - 3 g_s^2  |F_{12-p}^{(3)}|^2  \ .\nn
\eea
A de Sitter solution would then always have $\del_{\rho}^2 V|_0 > 0$, irrespectively of $\lambda$.\footnote{This result may look surprising, based on the intuition of Minkowski solutions on Calabi-Yau manifolds, where the volume, given in terms of the K\"ahler moduli, is not stabilized. This tension is made more explicit by rewriting \eqref{potrhorho} as follows using \eqref{eqdelrhoV}
\beq
\hspace{-0.3in} \ \rho^2 \del^2_\rho \tV|_0 =  3 |H|^2  - g_s  \frac{1}{4} (p-4) (p-6) \frac{T_{10}}{p+1} + g_s^2 \bigg(  \sum_{q=0}^{4} \frac{1}{2} (3-q)(4-q) |F_q|^2  - |F_5|^2 - 6 |F_6|^2 \bigg) \ . \label{potrhorho2}
\eeq
In the seminal Minkowski solution on a Calabi-Yau manifold of \cite{Giddings:2001yu} with $A=0$ and constant dilaton, the only non-trivial contributions are $|H|^2 = g_s^2 |F_3|^2  = g_s \frac{T_{10}}{p+1} >0$ with $p=3$ (see \cite{Andriot:2016ufg} for the appropriate conventions). We deduce in that case that  $\del^2_\rho V|_0 > 0$. This puzzle is solved by looking at the definition of moduli. Leaving $\sigma$ aside for the discussion, we are here considering $V(\rho,\tau)$, but $\tau$ actually depends on $\rho$ and the more fundamental dilaton fluctuation $\delta \phi$. What should be studied is thus $\tfrac{\d^2 V}{\d \rho^2} = \del^2_\rho V + 2 \del_\rho \del_\tau V \del_\rho \tau + \del_\tau V \del^2_\rho \tau + \del^2_\tau V (\del_\rho \tau)^2 $, which should vanish for that solution. A simpler way to proceed is to trade $\tau$ for $\delta \phi$ and $\rho$ in $V$: then, one verifies that on this precise solution, the dependence on $\rho$ disappears from $V$, as initially expected. Similar remarks are made in \cite{Hertzberg:2007wc} (see footnotes 4 and 9) on the different conventions for the volume modulus, and the authors switch to $V(\rho, \phi)$ when considering the solution of \cite{Giddings:2001yu}. Regarding de Sitter solutions, a related point is the fact the conditions $\del_{\varphi_i}^2 V|_0 > 0$ are only necessary to stability.} The same holds for $\sigma$ as long as $\lambda >0$, since we show
\bea
p=6:\quad & {\cal R}_4 + \frac{1}{2} \ \tau \del_{\tau} \tV|_0  -\frac{7}{45} \ \sigma \del_{\sigma} \tV|_0 - \frac{8}{135} \ \left(\sigma^2 \del^2_{\sigma} \tV|_0 + \sigma \del_{\sigma} \tV|_0 \right)\label{combisigmastab}\\
& = -\frac{6}{5} |f^{{}_{||}}{}_{{}_{\bot} {}_{\bot}}|^2 -\frac{6}{5} |H^{(0)}|^2 + \frac{14}{15} \delta^{cd}   f^{b_{\bot}}{}_{a_{||}  c_{\bot}} f^{a_{||}}{}_{ b_{\bot} d_{\bot}}  \nn\\
p=5:\quad & {\cal R}_4 +\frac{1}{2} \ \tau \del_{\tau} \tV|_0   -\frac{1}{15} \ \rho \del_{\rho} \tV|_0 -\frac{1}{15} \ \sigma \del_{\sigma} \tV|_0  -\frac{1}{20} \ \left(\sigma^2 \del^2_{\sigma} \tV|_0 + \sigma \del_{\sigma} \tV|_0 \right)\nn\\
& =   -\frac{4}{5} |f^{{}_{||}}{}_{{}_{\bot} {}_{\bot}}|^2  -\frac{1}{10}  |H^{(0)}|^2  - \frac{1}{2} |H^{(2)}|^2  -\frac{1}{10} g_s^2 |F_1|^2 + \delta^{cd}   f^{b_{\bot}}{}_{a_{||}  c_{\bot}} f^{a_{||}}{}_{ b_{\bot} d_{\bot}} \nn\\
p=4:\quad & \frac{59}{2} {\cal R}_4 + \frac{67}{4} \ \tau \del_{\tau} \tV|_0  + 8 \ \rho \del_{\rho} \tV|_0   - \ \sigma \del_{\sigma} \tV|_0 -\frac{1}{2} \ \left(\sigma^2 \del^2_{\sigma} \tV|_0 + \sigma \del_{\sigma} \tV|_0 \right)\nn\\
& = -24 |H^{(2)}|^2  -\frac{1}{8} g_s \frac{T_{10}}{p+1} -10 g_s^2 |F_4|^2 + 18 \delta^{cd}   f^{b_{\bot}}{}_{a_{||}  c_{\bot}} f^{a_{||}}{}_{ b_{\bot} d_{\bot}} \ .\nn
\eea
Finally, we prove a similar result for $\tau$, as long as $0<\lambda <1/17$ for $p=6$ and $0 < \lambda < 1/10$ for $p=4,5$. Indeed, we show for $10 \lambda - 1 \neq 0$, with $p=4,5,6$ and using $F_{q\geq 7}=0$ and $F_q^{(n> p-3 )}=0$, that
\bea
& \frac{1- 10 \lambda}{2}\ {\cal R}_4 + \frac{3- 32 \lambda}{4} \ \tau \del_{\tau} \tV|_0  + \frac{6-p}{6} \ \rho \del_{\rho} \tV|_0  + \frac{1}{6} \ \sigma \del_{\sigma} \tV|_0 \\
&  - \lambda\ \tau^2 \del^2_{\tau} \tV|_0  + (1 - \lambda) \left|*_{\bot}H^{(0)} + \varepsilon_p g_s F_{k-2}^{(0)} \right|^2  + (1 - \lambda) \sum_{a_{||}} \left| *_{\bot}( \d e^{a_{||}})|_{\bot} - \varepsilon_p g_s\, \iota_{a_{||}} F_k^{(1)} \right|^2 \nn\\
= & -\lambda |H^{(0)}|^2  - (1- \lambda + 8 \lambda (p-6)) g_s^2  |F_{10-p}^{(2)}|^2 - (1 - 17 \lambda) g_s^2  |F_{12-p}^{(3)}|^2 \ .\nn
\eea
For the stated values of $\lambda$, the right-hand side is negative, hence the result $\del^2_{\tau} V|_0 >0$ for a de Sitter solution.

We conclude that in the range $0<\lambda <1$ where having a de Sitter solution is not excluded, there exists a smaller range, namely $0<\lambda <1/17$ for $p=6$, and $0 < \lambda < 1/10$ for $p=4,5$, where for all three scalars we obtain $\del_{\varphi_i}^2 V|_0 > 0$, a necessary condition for them to be stabilized. The presence of such a stability island was completely unexpected, and it certainly deserves more investigation.

\section{Summary and swampland discussion}\label{sec:ccl}

In this paper, we have studied the existence and stability of classical de Sitter solutions in type II supergravities, with parallel $D_p/O_p$ sources of single size $p$. Motivated by \cite{Danielsson:2011au}, we have introduced a third 4d scalar field $\sigma$ on top of the two standard ones $\rho$ and $\tau$, and determined explicitly the general 4d scalar potential $V(\rho,\tau,\sigma)$, given in \eqref{pot0N} or \eqref{potN}. As explained in the Introduction, we have shown that the 4d Einstein equation and the three extrema conditions $\del_{\varphi_i} V|_0 = 0$ are equivalent to the four 10d equations of motion used in \cite{Andriot:2016xvq}, therefore providing a 4d counterpart to this previous 10d approach. This result is obtained at least in the case of constant warp factor and dilaton, and it would be interesting to extend this equivalence more generally. The advantage of the 4d approach is the possibility of computing the second derivatives of $V$, from which one reads stability constraints. This framework however did not allow us to conclude in full generality with e.g.~a no-go theorem. Specifying to group manifolds with constant fluxes, we reached better results. We first excluded the existence of de Sitter solutions in a large region of parameter space, in \eqref{nogo1} and \eqref{nogo2}. Provided one uses a standard basis for the Lie algebra (see the end of Section \ref{sec:nogo}), we then proved the absence of a solution on nilmanifolds (as noticed with intersecting sources after a scan in \cite{Danielsson:2011au}, and with consequences on the 4d theory \cite{Andriot:2018wzk}), on group manifolds with semi-simple algebras, and on some solvmanifolds. In the small remaining parameter region, $0< \lambda <1$, we could not conclude, giving the remaining entries of Table \ref{tab:after}. But we then identified in Section \ref{sec:stab} a subregion, namely $0<\lambda <1/17$ for $p=6$ and $0 < \lambda < 1/10$ for $p=4,5$, coined the stability island, where we showed that $\del^2_{\varphi_i} V|_0 > 0$ for the three scalars, for any de Sitter solution. It would be worth exploring this island in more detail.

An initial motivation in considering $\sigma$ was the observation, in \cite{Danielsson:2012et}, that for all explicit de Sitter solutions of \cite{Danielsson:2011au}, the tachyon was argued to be present within the three scalars $\rho,\tau,\sigma$ (see also \cite{Junghans:2016uvg}). With the formalism developed here, we could then hope to show the presence of a systematic tachyon, and prove a general no-go theorem against classical stable de Sitter solutions. This is not the result that we obtained: the no-go theorems derived here were only on the existence of solutions, and the only conclusive statement concerning stability was about stable solutions, instead of tachyons. Of course, an important difference is that we restricted ourselves to parallel sources, while the solutions of \cite{Danielsson:2011au} have intersecting $O_6$. We should then naturally extend the formalism developed here to the case of intersecting sources, using the 10d equivalent approach \cite{Andriot:2017jhf}. However, this situation also leaves room to a different understanding of a systematic tachyon, as for instance the one proposed in the line of papers \cite{Covi:2008ea, Kallosh:2014oja, Junghans:2016uvg, Junghans:2016abx} where de Sitter solutions close to no-scale Minkowski solutions were analysed.\\

Our results are important in the context of the swampland discussion, presented in the Introduction. As shown with Table \ref{tab:before} and \ref{tab:after}, we have restricted further the set of possibilities to find a classical de Sitter solution with parallel sources. It would then be interesting to determine the coefficient $c$ of the criterion of \cite{Obied:2018sgi}, in the newly excluded cases, and check whether it is of order $1$ as argued there. Some coefficients appearing in the new no-go theorems \eqref{nogo1} and \eqref{nogo2} are rather of order $1/10$, so it would interesting to check the details. Doing so may also require to determine the kinetic terms for the three scalar fields, especially that of $\sigma$, something that was not needed in the present work. Finally, the fact we only obtained no-go theorems on the existence of solutions, not involving the stability, favors the criterion of \cite{Obied:2018sgi} with respect to the refined criterion proposed in \cite{Andriot:2018wzk}. It is even more surprising, given that we have a priori allowed for contributions of the second derivatives, with our method using the general linear combination \eqref{galmethod}. However, as argued previously, the role of the second derivatives (and the presence of tachyons) may become more prominent in the case of intersecting sources, so the criterion of \cite{Andriot:2018wzk} and related coefficients determinations of \cite{Garg:2018reu} could then become more relevant. Last but not least, this whole discussion assumes the scalar potential and 4d theory \eqref{S4dUN} to be a low energy effective theory of string theory, so that it can be used to test the criteria, but this assumption could be further discussed \cite{Andriot:2018wzk}. The equivalence \eqref{3eq} and \eqref{1eq} to some 10d equations of motion rather points towards a consistent truncation, which does not necessarily match a low energy truncation; it would be interesting to study this question further.

Regarding the swampland discussion, the role of the (sourced) flux Bianchi identity is particularly interesting. As mentioned in the Introduction and Section \ref{sec:gal}, the 10d Bianchi identity \eqref{BI2} or \eqref{BI} has no 4d interpretation and should be included by hand in 4d to complete the theory. As such, it can be viewed as a quantum gravity (here stringy) consistency condition or requirement. This goes precisely in the direction of the swampland idea and the criterion of \cite{Obied:2018sgi} as we now illustrate. Consider for instance the 4d theory obtained with an $O_3$ and only $\rho,\tau$, coupled to gravity, with the potential $V(\rho,\tau)$. As mentioned in Table \ref{tab:before}, and as one can verify explicitly using equations \eqref{eqR4}, \eqref{eqdeltauV}, \eqref{eqdelrhoV}, there exist from the 4d perspective de Sitter solutions without fluxes, with
\beq
{\cal R}_4 = -\frac{2}{3} {\cal R}_6 = g_s \frac{T_{10}}{4} \ ,
\eeq
where ${\cal R}_6$ and $T_{10}$ are just parameters in the 4d potential. We now consider the Bianchi identity: without fluxes, it implies that $T_{10}=0$, which effectively forbids having de Sitter solutions. It then looks as if the quantum gravity origin brings a constraint on the question of having de Sitter solution, in agreement with the criterion of \cite{Obied:2018sgi}. A similar illustration is provided through our no-go theorem \eqref{nogo1}: the corresponding expression \eqref{lincombi1} is schematically of the form
\beq
{\cal R}_4  + b\ \varphi \del_{\varphi} V + |\mbox{BPS}|^2 \leq 0  \ \Rightarrow \ {\cal R}_4  + b\ \varphi \del_{\varphi} V  \leq 0  \ ,
\eeq
where the inequality on the right is analogous to the criterion of \cite{Obied:2018sgi}. Proving the first inequality was only made possible thanks to the Bianchi identity: it provides an extra relation between various quantities, including the $|\mbox{BPS}|^2$, allowing to eventually get a definite sign and conclude on the absence of de Sitter solution. As illustrated here, the Bianchi identity, viewed as a quantum gravity requirement, may therefore be a key to justify better the de Sitter swampland criterion of \cite{Obied:2018sgi}.

\vspace{0.4in}

\subsection*{Acknowledgements}

I would like to thank B.~Basso, G.~Shiu, P.~Tourkine, J.~Troost, T.~Van Riet, M.~Walters and T.~Wrase for helpful and stimulating discussions.

\newpage

\begin{appendix}

\section{An alternative scalar field}\label{ap:sigma}

We consider in this appendix a different definition for the scalar field $\sigma$, as being the following fluctuation
\beq
\d s_6^2 = \rho \left( \sigma (\d s_{||}^2)^0 + (\d s_{\bot}^2)^0 \right) \ , \ e^{a_{||}}{}_{m} = \sqrt{\rho \sigma}\, (e^{a_{||}}{}_{m})^0 \ ,\ e^{a_{\bot}}{}_{m} = \sqrt{\rho}\, (e^{a_{\bot}}{}_{m})^0 \label{fluctmetric} \ .
\eeq
The rescaling to Einstein frame leads to the following definition of $\tau$
\beq
\phi = \phi^0 + \delta \phi \ ,\ \tau = e^{- \delta \phi} \rho^{\frac{3}{2}} \sigma^{\frac{p-3}{4}} \ . \label{fluctdil}
\eeq
These scalars fields can be related to those of the main text, proposed in \cite{Danielsson:2012et}, by a reshuffling of $\rho, \sigma$. The background value is still recovered for $\rho=\sigma = \tau =1$. The rescaling to Einstein frame and definition of the Planck mass as given in \eqref{EM} still hold, as well as the formal definition \eqref{potfirst} of the potential $V$. The presence of $\sigma$ in $\tau$ modifies the 4d action towards
\beq
\hspace{-0.1cm} {\cal S}= M_4^2 \int \d x^{4} \sqrt{|g_{4E}|} \left( {\cal R}_{4E} - \frac{1}{2} \tau^4 \rho^{3} \sigma^{\frac{p-3}{2}} g_s^2 \left( |F_4^4|^{2E} + \frac{1}{2}|H_4|^{2E} |f_5(\rho,\sigma)|^2 \right) - {\rm kin} - \frac{1}{M_4^2} U \right) \ ,\nn
\eeq
and the potential to
\beq
\frac{1}{M_4^2}\ V =  \frac{1}{M_4^2}\ U - \frac{1}{2} \tau^4 \rho^{3} \sigma^{\frac{p-3}{2}} g_s^2 \left( |F_6^0|^{2} + \frac{1}{2} |(*_6 F_5^0)(\rho,\sigma)|^2 \right) \ .
\eeq
We recompute the potential and obtain, in the simplified notations,
\bea
\frac{1}{M_4^2}\ V = & - \tau^{-2} \bigg( \rho^{-1} {\cal R}_6(\sigma) -\frac{1}{2} \rho^{-3} \sum_n \sigma^{-n} |H^{(n)}|^2 \bigg) \label{pot}\\
& - g_s \tau^{-3} \rho^{\frac{p-6}{2}} \sigma^{\frac{p-3}{4}} \frac{T_{10}}{p+1} \nn\\
& +\frac{1}{2} g_s^2 \bigg( \tau^{-4} \sum_{q=0}^{4} \rho^{3-q} \sum_n  \sigma^{\frac{p-3}{2}-n} |F_q^{(n)}|^2  - \tau^{4} \rho^3  \sigma^{\frac{p-3}{2}} |F_6|^2 \nn\\
& \phantom{+\frac{1}{2} g_s^2  \bigg(} + \frac{1}{2}  \sum_n  \sigma^{\frac{p-3}{2}-n} (\tau^{-4} \rho^{-2} |F_5^{(n)}|^2 - \tau^{4} \rho^2 |(*_6 F_5)^{(n)}|^2 ) \bigg) \ . \nn
\eea
Note we do not match the main text expression by setting $A=1$, $B=0$, because these values are not a solution to the constraint \eqref{ABeq}, and this constraint was used in the definition $\tau$. The Ricci scalar dependence on $\sigma$ is now given by
\bea
{\cal R}_6 (\sigma) = &\ \left( {\cal R}_{\bot} +  \delta^{ab} \del_{a_{\bot}}  f^{c_{\bot}}{}_{c_{\bot}b_{\bot}} + {\cal R}_{\bot}^{||} + |f^{{}_{||}}{}_{{}_{\bot} {}_{\bot}}|^2 \right)^0 + \sigma^{-1} \left( {\cal R}_{||} + \delta^{ab} \del_{a_{||}}  f^{c_{||}}{}_{c_{||}b_{||}}  +  {\cal R}_{||}^{\bot} +|f^{{}_{\bot}}{}_{{}_{||} {}_{||}}|^2 \right)^0 \nn\\
& - \frac{1}{2} \sigma^{-2} |f^{{}_{\bot} 0}{}_{{}_{||} {}_{||}}|^2 - \frac{1}{2} \sigma |f^{{}_{||} 0}{}_{{}_{\bot} {}_{\bot}}|^2 \ . \label{R6sigma}
\eea

The Einstein equation and the derivatives of $V$ with respect to $\rho, \tau$, evaluated in the background, are not modified by this alternative definition. We now compute the derivative with respect to the present $\sigma$
\bea
\frac{1}{M_4^2}\ \sigma \del_\sigma V|_0 = &\  {\cal R}_{||} + \delta^{ab} \del_{a_{||}}  f^{c_{||}}{}_{c_{||}b_{||}}  +  {\cal R}_{||}^{\bot} + \frac{1}{2} |f^{{}_{||}}{}_{{}_{\bot} {}_{\bot}}|^2 - \frac{1}{2} \sum_n  n |H^{(n)}|^2 \\
& - g_s \frac{p-3}{4} \frac{T_{10}}{p+1} \nn\\
& +\frac{1}{2} g_s^2 \bigg( \sum_{q=0}^{4} \sum_n  \big(\frac{p-3}{2}-n \big) |F_q^{(n)}|^2  - \frac{p-3}{2} |F_6|^2 \nn\\
& \phantom{+\frac{1}{2} g_s^2  \bigg(} + \frac{1}{2}  \sum_n  \big(\frac{p-3}{2}-n \big) ( |F_5^{(n)}|^2 - |(*_6 F_5)^{(n)}|^2 ) \bigg) \ . \nn
\eea
Setting that derivative to zero and using \eqref{R4T10F}, we deduce
\bea
(\del_\sigma V): \qquad 0 =&\ {\cal R}_{||} + \delta^{ab} \del_{a_{||}}  f^{c_{||}}{}_{c_{||}b_{||}}  +  {\cal R}_{||}^{\bot} + \frac{1}{2} |f^{{}_{||}}{}_{{}_{\bot} {}_{\bot}}|^2  - \frac{p-3}{4} ({\cal R}_4 + g_s^2 |F_5|^2 + 2 g_s^2 |F_6|^2) \nn\\
& -\frac{1}{2} \sum_n  n \bigg( |H^{(n)}|^2 +  g_s^2 \sum_{q=0}^{4} |F_q^{(n)}|^2 + \frac{g_s^2}{2}  ( |F_5^{(n)}|^2 - |(*_6 F_5)^{(n)}|^2 ) \bigg) \ . \label{eqdelsigV}
\eea
This matches precisely the $(\del_\sigma V)$ condition \eqref{eqdelsigVN} obtained with the other definition of $\sigma$, using the values \eqref{ABvalues} for $A, B$. This is not surprising since the two first derivatives $\del_\sigma V$ should be related via a linear combination with the first derivatives with respect to $\rho, \tau$. Setting all of those to zero eventually gives the same condition. This alternative definition of $\sigma$ thus does not bring new information.

We turn to the second derivative and obtain
\bea
\sigma^2 \del^2_\sigma \tV|_0 = & - 2 \left( {\cal R}_{||} + \delta^{ab} \del_{a_{||}}  f^{c_{||}}{}_{c_{||}b_{||}}  +  {\cal R}_{||}^{\bot} \right) + |f^{{}_{\bot}}{}_{{}_{||} {}_{||}}|^2  \label{potsigmasigma}\\
& + \frac{1}{2}  \sum_n n(n+1) |H^{(n)}|^2  - g_s  \frac{1}{16} (p-3) (p-7) \frac{T_{10}}{p+1} \nn\\
& +\frac{1}{2} g_s^2 \bigg(  \sum_{q=0}^{4} \sum_n  \left(\frac{p-3}{2}-n\right) \left(\frac{p-3}{2}-n-1\right) |F_q^{(n)}|^2 \nn\\
& \phantom{+\frac{1}{2} g_s^2 \bigg( }  + \frac{1}{2}  \sum_n  \left(\frac{p-3}{2}-n\right) \left(\frac{p-3}{2}-n-1\right) ( |F_5^{(n)}|^2 - |(*_6 F_5)^{(n)}|^2 ) \nn\\
& \phantom{+\frac{1}{2} g_s^2 \bigg( }  -  \frac{1}{4} (p-3) (p-5) |F_6|^2 \bigg) \ . \nn
\eea
We now face the same problems as described at the end of Section \ref{sec:equations}: we schematically have
\bea
&\del_\sigma V \sim {\cal R}_{||} +  {\cal R}_{||}^{\bot} + \tfrac{1}{2} |f^{{}_{||}}{}_{{}_{\bot} {}_{\bot}}|^2  \label{configsign}\\
& \del_\sigma^2 V \sim - ( {\cal R}_{||} +  {\cal R}_{||}^{\bot}) \ ,\nn
\eea
which does not help in concluding on a systematic tachyon, despite the different definition for $\sigma$.

\section{An example and further comments}\label{ap:ex}

In Section \ref{sec:nogo}, we identified a region in parameter space, $0 < \lambda < 1$, for which we were not able to conclude on the absence of de Sitter solutions, opening the door to the possibility of having solutions there. In this appendix, we illustrate this case with an example of group manifold that passes the constraints discussed at the end of Section \ref{sec:nogo}, and an appropriate field ansatz. We consider the solvmanifold whose algebra is given by $(f_1\, 23, - f_2\, 13, 0, f_4\, 56, 0, 0)$ with $f_1 f_2 >0$: it is the direct product of a three-dimensional solvmanifold with standard rotation fibration, with a three-dimensional nilmanifold. We work in type IIB supergravity with an $O_5$ along $e^1 \w e^4$. This gives the following structure constants and curvature terms
\bea
& f^{{}_{||}}{}_{{}_{\bot} {}_{\bot}}:\ f^1{}_{23}=-f_1 \ ,\ f^4{}_{56}=-f_4 \ ,\quad f^{{}_{\bot}}{}_{{}_{||} {}_{\bot}}:\ f^2{}_{13}=f_2 \ , \nn\\
& |f^{{}_{||}}{}_{{}_{\bot} {}_{\bot}}|^2 = f_1^2 + f_4^2 \ ,\ {\cal R}_{||}=0 \ ,\ {\cal R}_{||}^{\bot} = -\frac{1}{2} f_2^2 \ ,\ -\delta^{cd}   f^{b_{\bot}}{}_{a_{||}  c_{\bot}} f^{a_{||}}{}_{ b_{\bot} d_{\bot}} = f_1 f_2 \ ,\\
& 2{\cal R}_6= - f_2^2 - f_1^2 - f_4^2 + 2 f_1 f_2 \ .\nn
\eea
We recall that we use the basis where the internal metric is $\delta_{ab}$, so the radii are present in the structure constants, on top of their quantization. We then consider the following values of the structure constants
\beq
f_1=f_2=f\ ,\ f_4=x f\ , \ f,x \in \mathbb{R}^* \ \Rightarrow \lambda=\frac{1}{1+x^2}<1 \ .
\eeq
We now allow for the following fluxes, where $F_5=0$ and the others depend on constants $a,b,c,d$
\bea
& F_1=c\, (e^5 + e^6) \ ,\ H=bc\, e^2\w e^3 \w (e^6 -e^5) + 2cd\, e^3 \w e^5 \w e^6 \ ,\\
& F_3= a\, e^1 \w e^2 \w (e^6 -e^5) + b\, (f_1\, e^1\w e^5 \w e^6 + f_4\, e^2\w e^3 \w e^4 ) + d\, f_4\, e^3 \w e^4 \w (e^5 + e^6) \ .\nn
\eea
These constant fluxes only have the components allowed by the orientifold projection \eqref{Fkns}, with $H^{(2)}=0$. These fluxes verify their Bianchi identities and equations of motion
\bea
& \d H= 0 \ ,\ \d F_1 = 0 \ ,\ \d F_3 - H \w F_1 = \frac{T_{10}}{p+1} e^2 \w e^3 \w e^5 \w e^6 \ ,\ H\w F_3=0\ ,\\
& \d(*H) - F_1 \w *F_3=0 \ ,\ \d (*F_1) + H \w * F_3 = 0 \ ,\ \d(*F_3)=0 \ ,
\eea
where we set $g_s=1$ for simplicity, and we get
\beq
\frac{T_{10}}{p+1} = b \left( |f^{{}_{||}}{}_{{}_{\bot} {}_{\bot}}|^2 + |F_1|^2  \right) \ .
\eeq
From there, we tried to solve the following linear combinations of the four-dimensional Einstein trace \eqref{eqR4}, $\del_\tau V|_0=0$ \eqref{eqdeltauV} and  $\del_\rho V|_0=0$ \eqref{eqdelrhoV}:
\bea
\eqref{eqR4} + \frac{3}{4} \eqref{eqdeltauV} + \frac{1}{2} \eqref{eqdelrhoV}:&\ \, 2 {\cal R}_4= \frac{T_{10}}{p+1} - |H|^2 - |F_3|^2\\
\frac{1}{2}  \eqref{eqdeltauV} +  \eqref{eqdelrhoV}:&\ -2 {\cal R}_6= 2\frac{T_{10}}{p+1} - 2|H|^2 - |F_3|^2\\
-\frac{1}{4}  \eqref{eqdeltauV} + \frac{1}{2}\eqref{eqdelrhoV}:&\ -\frac{1}{2}\frac{T_{10}}{p+1} - \frac{1}{2}|H|^2 + \frac{1}{2} |F_3|^2 + |F_1|^2 = 0
\eea
as well as $\del_{\sigma} V|_0 =0$ from \eqref{delVsigmaN}, that becomes here
\bea
0 = & \ 2 {\cal R}_6 -6 \left( {\cal R}_{||} +  {\cal R}_{||}^{\bot} \right)  - 3 |f^{{}_{||}}{}_{{}_{\bot} {}_{\bot}}|^2  - 3 |H|^2 +4 \frac{T_{10}}{p+1} - |F_1|^2 \ .
\eea
We did not find a solution with ${\cal R}_4 > 0$. It is not clear at this stage whether there is a way to conclude negatively on the existence of de Sitter solutions in the case $0<\lambda <1$, only with the constraints ${\cal R}_4>0$, $\del_{\tau,\, \rho,\, \sigma} V|_0 =0$, and the $F_3$ Bianchi identity, as achieved for the other cases, or whether the additional information included here, namely the other fluxes Bianchi identities, and the flux equations of motion, play a role. This indicates in any case that there is room to constrain more the case $0<\lambda <1$.\\

We make here a final comment on the method of the linear combination \eqref{galmethod} used to exclude solutions. One may have the impression that requiring each line to have a definite sign is too strong. For instance, focusing only on two lines and two coefficients:
\beq
g_s \frac{T_{10}}{p+1} \ (-2a +3b_{\tau}) + g_s^2  |F_q|^2 (a-2b_{\tau} ) \ , \label{sum}
\eeq
we are looking with our method for a set of constants $a, b_{\tau}$ such that $-2a +3b_{\tau} \leq 0$ and $a-2b_{\tau} \leq 0$. One may argue that having the sum \eqref{sum} negative is enough to exclude de Sitter solutions. More precisely, one could think of building combinations of definite signs, such as $-g_s \frac{T_{10}}{p+1} + g_s^2  |F_q|^2 \sim - {\cal R}_4$ (forgetting about the sum on fluxes) according to \eqref{R4T10F}. Building this combination would rather require the signs $-2a +3b_{\tau} \leq 0$ and $a-2b_{\tau} \geq 0$, which is less constraining than the initial sign requirement. From this perspective, it is true that our method appears too constraining. However, any combination of definite sign would actually be built as a linear combination of the finite set of equations that we use in the left-hand side of \eqref{galmethod}: it is for instance the case of \eqref{R4T10F} built as ${\cal R}_4 + \frac{1}{M_4^2}\ \del_{\tau} V|_0$. Making such combinations of definite sign appear on the right-hand side is therefore not an extra freedom, but it is captured by the general linear combination in the left-hand side of \eqref{galmethod}. In other words, any combination of definite sign can be absorbed through a redefinition (with a shift) of the constants $a, b_{\tau}$, etc., such that the right-hand side is not changed in the end.

Another formulation is to say that each of the quantities appearing in the right-hand side ($T_{10}, |F_q|^2$, etc.) are independent quantities, and therefore the sign of each of their coefficients should independently be fixed (negatively) to claim on an exclusion of a general solution, general meaning where any of these independent quantities can appear. However, at first sight, one would think that these quantities are not independent but connected through some relations. But such relations are precisely those considered on the left-hand side of \eqref{galmethod}, and the freedom in choosing such relations is captured through the free coefficients $a, b_{\tau}$, etc. So in the end, we believe that there is no refinement of this method to exclude solutions.

\end{appendix}

\end{document}